\documentclass[a4paper,11pt]{article}
\pdfoutput=1

\usepackage{jheppub}
\usepackage{graphicx}
\usepackage{bm}
\usepackage{etoolbox}
\usepackage{booktabs}
\usepackage{array}
\usepackage{placeins}
\usepackage{float}
\usepackage[colorinlistoftodos]{todonotes}

\usepackage{makecell}
\usepackage{multirow}

\title{AI-assisted modeling and Bayesian inference of unpolarized quark transverse momentum distributions from Drell--Yan data}

\author[a,b,c]{Zhong-Bo Kang,}
\author[a,b]{Luke Sellers,}
\author[a,b]{Congyue Zhang,}
\author[a]{and Curtis Zhou}

\affiliation[a]{Department of Physics and Astronomy, University of California, Los Angeles, CA 90095, USA}
\affiliation[b]{Mani L. Bhaumik Institute for Theoretical Physics, University of California, Los Angeles, CA 90095, USA}
\affiliation[c]{Center for Frontiers in Nuclear Science, Stony Brook University, Stony Brook, NY 11794, USA}

\emailAdd{zkang@physics.ucla.edu}
\emailAdd{lsellers1@g.ucla.edu}
\emailAdd{maxzhang2002@g.ucla.edu}
\emailAdd{curtiszhou@g.ucla.edu}

\abstract{We present an extraction of unpolarized quark transverse-momentum-dependent parton distribution functions (TMD PDFs) from Drell--Yan data within a Bayesian inference framework, incorporating artificial intelligence at multiple stages of the analysis. Our analysis is performed at ${\rm N^3LO}$ in perturbative QCD combined with ${\rm N^4LL}$ resummation accuracy. We first employ an AI-driven iterative procedure to explore and rank candidate functional forms for the nonperturbative contributions to TMD PDFs at the initial scale, as well as for the Collins-Soper evolution kernel, using $\chi^2$ fits and physics constraints. To enable efficient Bayesian inference, we construct a surrogate model for TMD cross sections by training a machine-learning emulator over the parameter space, replacing computationally expensive repeated evaluations and allowing scalable sampling with an affine-invariant Markov Chain Monte Carlo (MCMC) ensemble. Using this framework, we perform a global analysis of Drell-Yan data from fixed-target, RHIC, and LHC experiments and extract TMD PDFs with quantified uncertainties. We compare the results with those obtained using the replica method and highlight differences in the resulting uncertainty estimates.}

\begin{document}
\maketitle

\section{Introduction} \label{sec:intro}
Transverse-momentum dependent parton distribution functions (TMD PDFs) provide a three-dimensional picture of the proton in momentum space, encoding both the longitudinal and transverse motion of partons~\cite{Accardi:2012qut}. They play a central role in processes characterized by multiple scales, in particular when a hard scale $Q$ is accompanied by a much smaller transverse momentum $q_T \ll Q$. In this regime, TMD factorization~\cite{Collins:2011zzd,Boussarie:2023izj} enables a systematic resummation of large logarithms of $q_T/Q$ and provides a framework for precision QCD phenomenology. With the increasing precision of Drell--Yan measurements from fixed-target, RHIC, Tevatron, and LHC experiments, together with the future Electron-Ion Collider program~\cite{AbdulKhalek:2021gbh,Alexandrou:2026cnj}, reliable extractions of TMD PDFs and their uncertainties have become increasingly important.

In recent years, TMD phenomenology has undergone substantial progress driven by both theoretical and methodological developments. On the theory side, perturbative ingredients entering TMD factorization are now known to very high accuracy, enabling phenomenological analyses at next-to-next-to-next-to-leading order (N$^3$LO) combined with next-to-next-to-next-to-next-to-leading logarithmic (N$^4$LL) resummation~\cite{Moos:2025sal}. On the phenomenology side, global analyses combining multiple processes, such as Drell--Yan and semi-inclusive deep-inelastic scattering (SIDIS), have enabled increasingly comprehensive determinations of TMD PDFs and fragmentation functions, including studies of flavor dependence and correlations with collinear PDFs~\cite{Bacchetta:2024qre}. More recently, simultaneous fits of TMD and collinear distributions have been performed~\cite{Barry:2025glq}, highlighting the deep connection between these frameworks and improving constraints on proton structure. In parallel, neural-network parametrizations have been introduced~\cite{Bacchetta:2025ara} to provide flexible, data-driven descriptions of the nonperturbative components, offering improved descriptions of experimental data compared to traditional functional forms. Furthermore, the combination of experimental data with lattice QCD inputs has recently been explored~\cite{Avkhadiev:2025wps}, particularly for the determination of the Collins--Soper kernel, offering new avenues to reduce theoretical uncertainties.

Despite this progress, a central challenge remains the reliable quantification of uncertainties in TMD extractions. Most existing analyses rely on the Monte Carlo replica method, in which pseudodata are generated according to the experimental covariance matrix and refitted to produce an ensemble of parameter sets. This approach is intuitive, widely used, and well established in hadronic phenomenology, and has proven successful in a broad range of TMD extractions~\cite{Moos:2025sal,Bacchetta:2024qre,Bacchetta:2024yzl,Barry:2025glq,Bacchetta:2025ara,Avkhadiev:2025wps,Echevarria:2020hpy,Alrashed:2021csd}.

Bayesian inference offers a complementary framework for uncertainty quantification, in which the full posterior probability distribution of the model parameters is constructed given the data and a specified prior. This approach has been successfully applied in several neighboring areas, including collinear PDF fitting \cite{Candido:2024bayesGP,Capel:2024qkm}, small-$x$ dipole amplitude determination~\cite{Casuga:2025etc,Mantysaari:2025ltq},  heavy-ion phenomenology \cite{PhysRevC.94.024907,JETSCAPE:2026chun},
neutron-star inference \cite{PhysRevD.99.084049,PhysRevD.111.034005}, and
joint analyses across multiple sectors \cite{Huth:2022}. Bayesian methods provide a direct probabilistic interpretation of uncertainties, make prior assumptions explicit, and naturally incorporate marginalization over nuisance parameters. Furthermore, the Bayesian framework readily accommodates the combination of heterogeneous information sources, such as experimental data and lattice QCD inputs, within a unified probabilistic description.

At a general level, the relation between frequentist resampling methods (such as the replica approach) and Bayesian inference has been extensively studied in the statistical literature \cite{Efron:1994bootstrap, Stuart_2010, Kass1990TheVO}. However, their quantitative comparison in realistic TMD extractions remains limited. Given the widespread use of both approaches, it is therefore natural to compare them directly when estimating uncertainties. Such comparisons go beyond a simple cross-check: they can help identify which aspects of an inference problem are most sensitive to the choice of methodology and clarify the regimes in which one framework may provide a more reliable description. In this sense, replica and Bayesian methods are best viewed not as competing alternatives, but as complementary probes of the uncertainty structure of TMD fits.

In this work, we present a global extraction of unpolarized quark TMD PDFs from Drell--Yan data within a Bayesian inference framework, incorporating artificial intelligence (AI) at multiple stages of the analysis. Our theoretical setup is based on TMD factorization at high perturbative accuracy, enabling a precise description of the perturbative components. To address the modeling of the nonperturbative sector, we employ an AI-driven workflow to systematically explore and rank candidate functional forms for the TMD PDFs and the Collins--Soper kernel, subject to theoretical constraints and phenomenological criteria. This procedure allows for a broader and less biased exploration of the ansatz space compared to traditional approaches.

A key technical ingredient of our analysis is the use of a machine-learning emulator to accelerate the evaluation of TMD cross sections. Bayesian inference requires repeated likelihood evaluations across a high-dimensional parameter space, which would be computationally prohibitive using direct theory calculations. By training a neural-network surrogate model on a set of exact predictions, we enable efficient posterior sampling using Markov Chain Monte Carlo methods while maintaining high accuracy.

Within the same theoretical and phenomenological framework, we also perform a replica-based analysis, allowing for a direct comparison between Bayesian and replica uncertainty estimates. This comparison provides insight into the robustness of the extracted uncertainties, the role of parameter correlations, and the extent to which different inference methodologies capture the underlying structure of the parameter space.

The paper is organized as follows. In Sec.~\ref{sec:formalism}, we review the theoretical framework for TMD factorization in Drell--Yan production. Sec.~\ref{sec:data_selection} describes the experimental datasets used in the analysis. In Sec.~\ref{sec:fit-setup}, we present the fit setup, including the AI-driven exploration of nonperturbative parameterizations and a brief review of the replica method as a non-Bayesian reference. The Bayesian inference framework and emulator construction are detailed in Sec.~\ref{sec:bayesian}. The results, including a comparison between Bayesian and replica uncertainties, are presented in Sec.~\ref{sec:results}. We conclude in Sec.~\ref{sec:conclusions}, and provide technical details on the uncertainty analysis, along with the posterior distributions, in the appendix.

\section{Theoretical formalism}
\label{sec:formalism}

\subsection{Drell--Yan process and kinematics}
\label{subsec:dy-kinematics}

We consider unpolarized neutral-current Drell--Yan production,
\begin{equation}
h_1(P_1)+h_2(P_2)\to \gamma^*/Z(q)\to \ell^+(l)+\ell^-(l')+X,
\label{eq:DY-process}
\end{equation}
where \(h_1\) and \(h_2\) are the incoming hadrons with momenta \(P_1^\mu\) and
\(P_2^\mu\), \(q^\mu\) is the momentum of the intermediate electroweak boson, \(l^\mu\)
and \(l'^\mu\) are the lepton momenta, and \(X\) denotes the unobserved hadronic final
state. The dilepton momentum is
\begin{equation}
q^\mu=l^\mu+l'^\mu,
\quad
Q^2\equiv q^2>0,
\label{eq:q-def}
\end{equation}
with \(Q\) the invariant mass of the lepton pair. Throughout we use the metric
\(g^{\mu\nu}=\mathrm{diag}(1,-1,-1,-1)\).

All kinematic variables are defined in the hadronic center-of-mass frame. We choose the
\(z\)-axis along the momentum of hadron \(h_1\). Neglecting hadron masses, the incoming
hadron momenta are
\begin{equation}
P_1^\mu=\frac{\sqrt{s}}{2}(1,0,0,1),
\quad
P_2^\mu=\frac{\sqrt{s}}{2}(1,0,0,-1),
\quad
s\equiv (P_1+P_2)^2 .
\label{eq:P1P2-com}
\end{equation}
It is then convenient to introduce the light-cone basis vectors and transverse projector,
\begin{equation}
n^\mu=\frac{1}{\sqrt{2}}(1,0,0,1),
\quad
\bar n^\mu=\frac{1}{\sqrt{2}}(1,0,0,-1),
\quad
g_T^{\mu\nu}=g^{\mu\nu}-n^\mu \bar n^\nu-\bar n^\mu n^\nu .
\label{eq:nbarn-def}
\end{equation}
Any four-vector \(v^\mu\) can then be decomposed as
\begin{equation}
v^\mu = v^+ n^\mu + v^- \bar n^\mu + v_T^\mu ,
\quad
v^+ \equiv v\cdot \bar n = \frac{v^0+v^3}{\sqrt{2}},
\quad
v^- \equiv v\cdot n = \frac{v^0-v^3}{\sqrt{2}},
\label{eq:v-decomp}
\end{equation}
with the transverse component given by
\begin{equation}
v_T^\mu = g_T^{\mu\nu} v_\nu,
\quad
v_T \equiv \sqrt{-v_T^\mu v_{T\mu}}.
\label{eq:lc-components}
\end{equation}

Neglecting lepton masses, the leptonic pseudorapidities $\eta$ and the dilepton rapidity $y$ are defined by
\begin{equation}
\eta_l = \frac{1}{2}\ln\frac{l^+}{l^-},
\quad
\eta_{l'} = \frac{1}{2}\ln\frac{l'^+}{l'^-},
\quad
y \equiv \frac{1}{2}\ln\frac{q^+}{q^-}.
\label{eq:eta-def}
\end{equation}
We denote by \(k_1^\mu\) and \(k_2^\mu\) the
momenta of the partons inside hadrons \(h_1\) and \(h_2\),
respectively. Their momenta are decomposed as
\begin{equation}
k_1^\mu = x_1 P_1^\mu + k_1^- \bar n^\mu + k_{1T}^\mu,
\quad
k_2^\mu = x_2 P_2^\mu + k_2^+ n^\mu + k_{2T}^\mu,
\label{eq:parton-momenta}
\end{equation}
where \(x_1\) and \(x_2\) are the longitudinal momentum fractions, while
\(k_{1T}^\mu\) and \(k_{2T}^\mu\) denote the transverse momenta of the two incoming
partons. At leading power in \(q_T/Q\), the partonic kinematics implies
\begin{equation}
q^+ = x_1 P_1^+ + \mathcal O\!\left(\frac{q_T^2}{Q}\right),
\quad
q^- = x_2 P_2^- + \mathcal O\!\left(\frac{q_T^2}{Q}\right),
\quad
q_T^\mu = k_{1T}^\mu + k_{2T}^\mu.
\label{eq:q-components}
\end{equation}
Accordingly, the partonic momentum fractions satisfy
\begin{equation}
x_1 
= \frac{Q}{\sqrt{s}}\,e^y + \mathcal O\!\left(\frac{q_T^2}{Q^2}\right),
\quad
x_2 
= \frac{Q}{\sqrt{s}}\,e^{-y} + \mathcal O\!\left(\frac{q_T^2}{Q^2}\right),
\label{eq:x1x2-def}
\end{equation}
so that
\begin{equation}
x_1x_2s = Q^2 + \mathcal O(q_T^2).
\label{eq:x1x2-lp-relation}
\end{equation}

\subsection{TMD factorization of the Drell--Yan cross section}
\label{subsec:lowqt-factorization}

In the low-$q_T$ region, the neutral-current Drell--Yan cross section with fiducial cuts
imposed on the final-state leptons can be written in the following form~\cite{Collins:2011zzd,Boussarie:2023izj}
\begin{equation}
\frac{\mathrm d\sigma}{\mathrm dq_T\,\mathrm dQ\,\mathrm dy}
=
\frac{16\pi^2\,\alpha_{\rm em}^2(Q)}{9\,Q\,s}\,q_T\,
H(Q,\mu_f)\,
\mathcal P\,
\sum_q c_q(Q)\,
\mathcal W_q(q_T,y,Q;\mu_f,\zeta_f)
+
\mathcal O\!\left(\frac{q_T^2}{Q^2}\right).
\label{eq:dy_lowqt_kt}
\end{equation}
Here $q \in \{ u,d,s,c,b, \bar u,\bar d,\bar s,\bar c,\bar b\}$, \(\mu_f\) and \(\zeta_f\) are the renormalization and rapidity scales, respectively,
and \(H(Q,\mu_f)\) is the hard function.
The electroweak coefficient for quark flavor \(q\) is
\begin{equation}
c_q(Q)
=
e_q^2
-2\chi_1(Q)\,e_qV_qV_\ell
+\bigl(V_\ell^2+A_\ell^2\bigr)\chi_2(Q)\,\bigl(V_q^2+A_q^2\bigr),
\label{eq:ew_factor}
\end{equation}
with
\begin{equation}
\chi_1(Q)
=
\frac{1}{4s_W^2c_W^2}\,
\frac{Q^2(Q^2-m_Z^2)}
{(Q^2-m_Z^2)^2+(m_Z\Gamma_Z)^2},
\qquad
\chi_2(Q)
=
\frac{1}{16s_W^4c_W^4}\,
\frac{Q^4}
{(Q^2-m_Z^2)^2+(m_Z\Gamma_Z)^2},
\label{eq:chi_factors}
\end{equation}
and
\begin{equation}
V_f=T_f^3-2e_fs_W^2,
\qquad
A_f=T_f^3,
\qquad
f \in \{q,\ell\}.
\label{eq:va_couplings}
\end{equation}
Here \(e_f\) denotes the electric charge in units of the positron charge,
\(T_f^3\) is the third component of weak isospin, \(s_W\equiv \sin\theta_W\),
\(c_W\equiv \cos\theta_W\), and \(m_Z\) and \(\Gamma_Z\) are the mass and width of the
\(Z\) boson, respectively.

To incorporate the leptonic phase space and fiducial cuts into the low-$q_T$
factorization formula, we define the phase-space reduction factor
\(\mathcal P\)~\cite{Scimemi:2017etj,Scimemi:2019cmh,Piloneta:2024aac,Neumann:2022lft} by
\begin{equation}
\mathcal P
=
\frac{\displaystyle
\int d\Phi_{\ell\ell}(q)\,
\Theta_{\rm cuts}(l,l')\,
(-g_T^{\mu\nu})L_{\mu\nu}(l,l')}
{\displaystyle
\int d\Phi_{\ell\ell}(q)\,
(-g_T^{\mu\nu})L_{\mu\nu}(l,l')}\, .
\label{eq:fiducial_factor}
\end{equation}
Here \(d\Phi_{\ell\ell}(q)\) is the two-body leptonic phase-space measure at fixed
dilepton momentum \(q^\mu\), and \(L_{\mu\nu}(l,l')\) is the standard Born leptonic tensor.
The step function \(\Theta_{\rm cuts}(l,l')\) implements the experimental acceptance on
the two final-state leptons, namely
\begin{equation}
l_T>l_{T,\min}^{(1)},
\qquad
l_T'>l_{T,\min}^{(2)},
\qquad
\eta_{\min}<\eta_l,\eta_{l'}<\eta_{\max}.
\label{eq:cut_conditions}
\end{equation}
In the inclusive limit one has \(\Theta_{\rm cuts}=1\), so that \(\mathcal P=1\).
The explicit dependence of \(\mathcal P\) on the cut parameters is suppressed
for brevity. During the fit, \(\mathcal P\) is evaluated using the numerical implementation
of fiducial cuts in the \texttt{arTeMiDe} framework
\cite{Vladimirov:2016dll,Vladimirov:2017ksc}, and collinear PDF are accessed through LHAPDF \cite{Buckley:2014ana}. Apart from these two, all other components of the cross section calculation  are implemented in an original code.

Within TMD factorization at leading power, the structure function \(\mathcal W_q\) is given by the standard transverse-momentum convolution of TMD PDFs~\cite{Collins:2011zzd,Boussarie:2023izj},
\begin{equation}
\begin{aligned}
\mathcal W_q(q_T,y,Q;\mu_f,\zeta_f)
={}&
\int d^2\bm k_{1T}\,d^2\bm k_{2T}\,
\delta^{(2)}\!\bigl(\bm q_T-\bm k_{1T}-\bm k_{2T}\bigr)
\\
&\times
F_q^{\,h_1}(x_1,\bm k_{1T};\mu_f,\zeta_f)\,
F_{\bar q}^{\,h_2}(x_2,\bm k_{2T};\mu_f,\zeta_f).
\end{aligned}
\label{eq:dy_tmd_convolution}
\end{equation}
Here and below, uppercase \(F\) denotes TMD PDFs, while lowercase \(f\) is reserved for
ordinary collinear PDFs. We define the TMD PDFs
in transverse coordinate $\bm b$ space as~\cite{Collins:1984kg,Collins:2011zzd,Becher:2010tm,Echevarria:2011epo}
\begin{equation}
\widetilde F_q^{\,h}(x,b;\mu_f,\zeta_f)
=
\int d^2\bm k_T\,
e^{-i\bm b\cdot\bm k_T}\,
F_q^{\,h}(x,\bm k_T;\mu_f,\zeta_f)\,,
\label{eq:tmd_fourier_transform}
\end{equation}
where $\bm b$ is Fourier-conjugate to the transverse momentum $\bm k_T$. 
The corresponding coordinate-space factorization then
becomes~\cite{Collins:1984kg,Collins:2011zzd,Becher:2010tm,Echevarria:2011epo,Scimemi:2017etj}
\begin{equation}
\begin{aligned}
\frac{\mathrm d\sigma}{\mathrm dq_T\,\mathrm dQ\,\mathrm dy}
= &
\frac{16\pi^2\,\alpha_{\rm em}^2(Q)}{9\,Q\,s}\,q_T\,
H(Q,\mu_f)\,
\mathcal P\,
\int_0^\infty \frac{db\,b}{2\pi}\,
J_0(b \, q_T)\,
\\
\times
&
\sum_q c_q(Q)
\widetilde F_q^{\,h_1}(x_1,b;\mu_f,\zeta_f)\,
\widetilde F_{\bar q}^{\,h_2}(x_2,b;\mu_f,\zeta_f)
+
\mathcal O\!\left(\frac{q_T^2}{Q^2}\right).
\label{eq:dy_lowqt_bspace}
\end{aligned}
\end{equation}

\subsection{Matching and evolution of TMDs}
\label{subsec:matching-evolution}

We write the quark TMD PDF entering the factorized cross section
as~\cite{Collins:2011zzd,Echevarria:2016scs,Gehrmann:2012ze,Gehrmann:2014yya}
\begin{equation}
\widetilde F_q^{\,h}(x,b;\mu,\zeta)
=
\left[
\sum_i
\int_x^1\frac{\mathrm d\xi}{\xi}\,
C_{q\leftarrow i}(\xi,b_\ast(b);\mu,\zeta)\,
f_i^{\,h}\!\left(\frac{x}{\xi},\mu\right)
\right]
S_{\rm NP}(x,b).
\label{eq:tmd_matching}
\end{equation}
Here, $i \in \{ u,d,s,c,b,g, \bar u,\bar d,\bar s,\bar c,\bar b\}$, and the matching coefficients \(C_{q\leftarrow i}\) encode the perturbative matching onto
collinear PDFs at small $b\ll 1/\Lambda_{\rm QCD}$, while \(S_{\rm NP}(x,b)\) parameterizes the nonperturbative large-\(b\) correction. The functional form for the implemented $b_\ast$ prescription is discussed in sec.~\ref{subsec:phenomenological-realization}.

Its scale dependence is governed
by the evolution equations~\cite{Collins:1981uk,Collins:2011zzd,Echevarria:2016scs,Scimemi:2017etj,Scimemi:2019cmh,Vladimirov:2020umg,Boussarie:2023izj}
\begin{equation}
\begin{aligned}
\frac{\partial}{\partial\ln\mu}\,
\ln \widetilde F_q^{\,h}(x,b;\mu,\zeta)
&=
\gamma_F(\mu,\zeta),
\\
\frac{\partial}{\partial\ln\zeta}\,
\ln \widetilde F_q^{\,h}(x,b;\mu,\zeta)
&=
-\mathcal D(b,\mu),
\end{aligned}
\label{eq:tmd_evolution_eqs}
\end{equation}
where \(\gamma_F\) is the anomalous dimension governing the \(\mu\)-dependence of the
TMD PDF, and \(\mathcal D\) is the Collins--Soper (CS) kernel.

The evolution of \(\gamma_F\) and \(\mathcal D\) is in turn determined
by~\cite{Collins:1981uk,Collins:2011zzd,Korchemsky:1985xj,Vladimirov:2020umg}
\begin{equation}
\begin{aligned}
\frac{\partial}{\partial\ln\zeta}\,
\gamma_F(\mu,\zeta)
&=
-\Gamma_{\rm cusp}(\mu),
\\
\frac{\partial}{\partial\ln\mu}\,
\mathcal D(b,\mu)
&=
\Gamma_{\rm cusp}(\mu),
\end{aligned}
\label{eq:kernel_evolution_eqs}
\end{equation}
with \(\Gamma_{\rm cusp}\) the cusp anomalous dimension. Consistency of the
\(\mu\)- and \(\zeta\)-evolution then
implies~\cite{Becher:2014oda,Echevarria:2016scs}
\begin{equation}
\gamma_F(\mu,\zeta)
=
\Gamma_{\rm cusp}(\mu)\ln\frac{\mu^2}{\zeta}
-\gamma_V(\mu),
\label{eq:gammaF_def}
\end{equation}
where \(\gamma_V\) is the non-cusp anomalous dimension.

For the phenomenological realization used below, the Collins--Soper kernel is
parameterized as~\cite{Scimemi:2017etj,Scimemi:2019cmh,Vladimirov:2016dll,Li:2016ctv,Moult:2022xzt,Duhr:2022yyp}
\begin{equation}
\mathcal{D}(b,\mu)
=
\mathcal{D}^{\rm pert}\!\left(b_\ast^{\rm CS}(b),\mu_\ast^{\rm CS}(b)\right)
+
\int_{\mu_\ast^{\rm CS}(b)}^{\mu}\frac{\mathrm{d}\mu'}{\mu'}\,
\Gamma_{\rm cusp}(\mu')
+
D_{\rm NP}(b),
\label{eq:cs_kernel_split}
\end{equation}
where \(\mathcal D^{\rm pert}\) denotes the perturbative boundary term of the Collins--Soper
kernel, and \(D_{\rm NP}(b)\) parameterizes the nonperturbative large-\(b\)
contribution. The profile choices for \(b_\ast^{\rm CS}(b)\),
\(\mu_\ast^{\rm CS}(b)\), and \(D_{\rm NP}(b)\) will be specified in
sec.~\ref{subsec:phenomenological-realization}.

We perform the matching at the initial scales \((\mu_i(b),\zeta_i(b))\), and then evolve the
TMD PDF to the final scales \((\mu_f,\zeta_f)\) through the evolution kernel
\(R(b;\mu_i(b),\zeta_i(b)\to\mu_f,\zeta_f)\)~\cite{Collins:1981uk,Collins:2011zzd,Echevarria:2016scs,Scimemi:2017etj,Scimemi:2019cmh,Vladimirov:2020umg,Boussarie:2023izj},
\begin{equation}
R(b;\mu_i,\zeta_i\to\mu_f,\zeta_f)
=
\exp\!\left[
\int_{\mu_i}^{\mu_f}\frac{\mathrm d\mu}{\mu}\,
\gamma_F(\mu,\zeta_f)
\right]
\left(\frac{\zeta_f}{\zeta_i}\right)^{-\mathcal D(b,\mu_i)}.
\label{eq:tmd_evolution_kernel}
\end{equation}
The matched-and-evolved TMD PDF then becomes~\cite{Collins:2011zzd,Echevarria:2016scs,Scimemi:2017etj,Scimemi:2019cmh}
\begin{equation}
\begin{aligned}
\widetilde F_q^{\,h}(x,b;\mu_f,\zeta_f)
= &
\left[
\sum_i
\int_x^1\frac{\mathrm d\xi}{\xi}\,
C_{q\leftarrow i}(\xi,b_\ast(b);\mu_i(b),\zeta_i(b))\,
f_i^{\,h}\!\left(\frac{x}{\xi},\mu_i\right)
\right]
\\
&
\times
R(b;\mu_i(b),\zeta_i(b)\to\mu_f,\zeta_f)\,
S_{\rm NP}(x,b).
\label{eq:tmd_evolved}
\end{aligned}
\end{equation}

\begin{table}[t]
    \centering
    \small
    \setlength{\tabcolsep}{10pt}
    \renewcommand{\arraystretch}{0.1}
    \begin{tabular}{ccccc}
        \toprule
        $H$ & $C_{q\leftarrow i}$ & $\gamma_V$ & $\mathcal D^{\rm pert}$ & $\Gamma_{\rm cusp}$ \\
        \midrule
        $ \alpha_s^3$ 
        & $\alpha_s^3$ 
        & $\alpha_s^4$ 
        & $\alpha_s^4$
        & $\alpha_s^{5}$ \\
        \bottomrule
    \end{tabular}
    \caption{Highest accuracy for perturbative ingredients entering the TMD factorization used in this work. This corresponds to nominal N\textsuperscript{3}LO + N\textsuperscript{4}LL accuracy.}
    \label{tab:perturbative_accuracy}
\end{table}

Before introducing the profile-based scales \((\mu_i,\zeta_i,\mu_f,\zeta_f)\) in
sec.~\ref{subsec:phenomenological-realization}, table~\ref{tab:perturbative_accuracy}
summarizes the perturbative ingredients entering the matched and evolved
TMD PDF in eq.~\eqref{eq:tmd_evolved}. We use the standard neutral-current Drell--Yan hard factor \(H\),
derived from the quark form factor, together with the small-\(b\)
operator-product-expansion matching coefficients \(C_{q\leftarrow i}\),
both through N\textsuperscript{3}LO~\cite{Collins:1984kg,Collins:2011zzd,Becher:2010tm,Echevarria:2011epo,Moch:2005tm,Gehrmann:2010ue,Echevarria:2016scs,Gehrmann:2012ze,Gehrmann:2014yya,Ebert:2020yqt,Luo:2020epw}.
The evolution kernel is implemented with the quark non-cusp anomalous
dimension \(\gamma_V\) and the perturbative Collins--Soper boundary term
\(\mathcal D^{\rm pert}\) through 4 loops~\cite{Collins:1981uk,Collins:2011zzd,Korchemsky:1985xj,Scimemi:2017etj,Agarwal:2021zft,Echevarria:2015byo,Vladimirov:2016dll,Li:2016ctv,Moult:2022xzt,Duhr:2022yyp}, and the cusp anomalous dimension to 5 loops
~\cite{Korchemsky:1985xj,Collins:2011zzd,Scimemi:2017etj,Henn:2019swt,Herzog:2018kwj,Moos:2023yfa,Moos:2025sal}. The collinear PDF and running of $\alpha_s$ are from the MSHT20aN\textsuperscript{3}LO sets~\cite{Bailey:2020ooq,McGowan:2022nag}.

\subsection{Phenomenological realization}
\label{subsec:phenomenological-realization}

We now specify the profile choices and fitted nonperturbative functions.
For the perturbative matching, we employ a $b_\ast$ prescription~\cite{Collins:1984kg} to interpolate between the small-$b$ and large-$b$ regions. In this work, we adopt the following functional form:
\begin{equation}
\begin{aligned}
b_\ast(b)
=
\frac{b}{\left[1+\left(b/b_{\rm max}\right)^4\right]^{1/4}},
\quad
\mu_\ast(b)
=
\frac{b_0}{b_\ast(b)},
\end{aligned}
\label{eq:bstar_profile}
\end{equation}
with $b_0=2e^{-\gamma_E}$ and $b_{\rm max}=b_0\,\mathrm{GeV}^{-1}$. This prescription preserves the perturbative small-$b$ region, where $b_\ast\simeq b$, while smoothly freezing the matching scale at large $b$. In particular, one has $b_\ast(b)\to b_{\rm max}$ as $b\to\infty$, so that $\mu_\ast(b)\to b_0/b_{\rm max}=1\,\mathrm{GeV}$.

For the perturbative Collins--Soper kernel, we instead use a separate regulating profile~\cite{Scimemi:2017etj,Scimemi:2019cmh,Moos:2025sal},
\begin{equation}
b_\ast^{\rm CS}(b)=\frac{b}{\sqrt{1+\left(b/b_{\rm max}\right)^2}},
\quad
\mu_\ast^{\rm CS}(b)=\frac{b_0}{b_\ast^{\rm CS}(b)}.
\label{eq:bstar_cs_profile}
\end{equation}
The perturbative scales are chosen as
\begin{equation}
\mu_i(b)=\mu_\ast(b),
\quad
\zeta_i(b)=\mu_\ast^2(b),
\quad
\mu_f=Q,
\quad
\zeta_f=Q^2.
\label{eq:phenomenological_scales}
\end{equation}

The $b_*$ prescription provides a smooth interpolation between the perturbative small-$b$ region and the nonperturbative large-$b$ regime. As this interpolation is not unique, different choices may be adopted for different components of the factorization formula. We therefore employ distinct $b_*$ prescriptions for the perturbative matching and the Collins--Soper kernel, allowing for independent control of the matching scale and the rapidity evolution, which probe different aspects of the underlying dynamics.

The final ansatz for the nonperturbative sector was selected through the AI-agent-driven exploration described in Sec.~\ref{subsec:agents}. For the nonperturbative Sudakov factor entering the TMD PDF in eq.~\eqref{eq:tmd_evolved}, we adopt a functional form inspired by Refs.~\cite{Moos:2023yfa,Moos:2025sal}:
\begin{equation}
S_{\rm NP}(x,b)=\operatorname{sech}\!\bigl[\mathcal{S}(x)\, b\bigr],
\label{eq:snp_mu}
\end{equation}
where the $x$-dependent shape function, generated from an AI-driven exploration for nonperturbative parameterizations, is parameterized as
\begin{equation}
\mathcal{S}(x)
=
\lambda_1\ln x
+\lambda_2(1-x)
+\lambda_3x(1-x)
+A_{\rm NP}\exp\!\left[
-\frac{(\ln x-\ln x_0)^2}{2\sigma_x^2}
\right].
\label{eq:xi_np}
\end{equation}
The nonperturbative function $S_{\rm NP}(x,b)$ therefore obeys
$S_{\rm NP}(x,b)=1-\tfrac12\,\mathcal{S}^2(x)b^2+\mathcal O(b^4)$ as $b\to0$,
while $S_{\rm NP}(x,b)\to0$ as $b\to\infty$.
For the nonperturbative Collins--Soper kernel, we adopt the ansatz obtained from the AI-driven exploration:
\begin{equation}
\begin{aligned}
D_{\rm NP}(b)
&=
b\,b_{\rm sat}(b)
\left[
c_0+c_1\ln\!\left(\frac{b_{\rm sat}(b)}{B_{\rm NP}}\right)
\right],
\quad
b_{\rm sat}(b)
=
\frac{b}{\sqrt{1+\left(b/B_{\rm NP}\right)^2}}\,.
\end{aligned}
\label{eq:dnp_model}
\end{equation}
This form is the same as that used in the ART TMD extractions~\cite{Moos:2023yfa,Moos:2025sal}.
The fit therefore contains 9 nonperturbative fit parameters:
\begin{equation}
\{c_0, c_1, B_{\rm NP}, \lambda_1, \lambda_2, \lambda_3, A_{\rm NP}, x_0, \sigma_x\}.
\label{eq:NP_parameters}
\end{equation}

\section{Data selection}
\label{sec:data_selection}

\begin{table}[t]
    \centering
    \footnotesize
    \setlength{\tabcolsep}{3pt}
    \renewcommand{\arraystretch}{1.08}
    \begin{tabular}{|c|c|c|c|c|c|c|c|}
        \hline
        Dataset & Process & Observable & \(\sqrt{s}\) [GeV] & \makecell[c]{Fiducial cut} & \(Q\) [GeV] & \(y\) & \(N_{\rm pts}\) \\
        \hline
        \makecell[c]{STAR~\cite{STAR:2019prelim}}
        & p-p
        & \(\mathrm d\sigma/\mathrm dq_T\)
        & 510
        & \makecell[c]{\(p_T>25\,\mathrm{GeV}\)\\\(|\eta|<1\)}
        & 73--114
        & \(|y|<1\)
        & 7 \\
        \hline
        \makecell[c]{CDF Run I~\cite{CDF:1999bpw}}
        & p-\(\bar{\mathrm p}\)
        & \(\mathrm d\sigma/\mathrm dq_T\)
        & 1800
        & Inclusive
        & 66--116
        & Inclusive
        & 25 \\
        \hline
        \makecell[c]{CDF Run II~\cite{CDF:2012brb}}
        & p-\(\bar{\mathrm p}\)
        & \(\mathrm d\sigma/\mathrm dq_T\)
        & 1960
        & Inclusive
        & 66--116
        & Inclusive
        & 26 \\
        \hline
        \makecell[c]{D0 Run I~\cite{D0:1999jba}}
        & p-\(\bar{\mathrm p}\)
        & \(\mathrm d\sigma/\mathrm dq_T\)
        & 1800
        & Inclusive
        & 75--105
        & Inclusive
        & 12 \\
        \hline
        \makecell[c]{D0 Run II~\cite{D0:2007lmg}}
        & p-\(\bar{\mathrm p}\)
        & \(\sigma^{-1}\,\mathrm d\sigma/\mathrm dq_T\)
        & 1960
        & Inclusive
        & 70--110
        & Inclusive
        & 5 \\
        \hline
        \makecell[c]{D0 Run II (\(\mu^+\mu^-\))~\cite{D0:2010dbl}}
        & p-\(\bar{\mathrm p}\)
        & \(\sigma^{-1}\,\mathrm d\sigma/\mathrm dq_T\)
        & 1960
        & \makecell[c]{\(p_T>15\,\mathrm{GeV}\)\\\(|\eta|<1.7\)}
        & 65--115
        & Inclusive
        & 3 \\
        \hline
        \makecell[c]{ATLAS 7 TeV~\cite{ATLAS:2014alx}}
        & p-p
        & \(\sigma^{-1}\,\mathrm d\sigma/\mathrm dq_T\)
        & 7000
        & \makecell[c]{\(p_T>20\,\mathrm{GeV}\)\\\(|\eta|<2.4\)}
        & 66--116
        & \makecell[c]{\(|y|<1\)\\\(1<|y|<2\)\\\(2<|y|<2.4\)}
        & \makecell[c]{6\\6\\6} \\
        \hline
        \multirow[c]{8}{*}{\makecell[c]{ATLAS 8 TeV~\cite{ATLAS:2015iiu}}}
        & \multirow[c]{8}{*}{p-p}
        & \multirow[c]{8}{*}{\(\mathrm d\sigma/\mathrm dq_T\)}
        & \multirow[c]{8}{*}{8000}
        & \multirow[c]{8}{*}{\makecell[c]{\(p_T>20\,\mathrm{GeV}\)\\\(|\eta|<2.4\)}}
        & \multirow[c]{6}{*}{66--116}
        & \(|y|<0.4\) & 6 \\
        & & & & & & \(0.4<|y|<0.8\) & 6 \\
        & & & & & & \(0.8<|y|<1.2\) & 6 \\
        & & & & & & \(1.2<|y|<1.6\) & 6 \\
        & & & & & & \(1.6<|y|<2.0\) & 6 \\
        & & & & & & \(2.0<|y|<2.4\) & 6 \\
        \cline{6-7}
        & & & & & \multirow[c]{2}{*}{\makecell[c]{46--66\\116--150}}
        & \multirow[c]{2}{*}{\(|y|<2.4\)} & 4 \\
        & & & & & & & 8 \\
        \hline
        \makecell[c]{CMS 7 TeV~\cite{CMS:2011wyd}}
        & p-p
        & \(\sigma^{-1}\,\mathrm d\sigma/\mathrm dq_T\)
        & 7000
        & \makecell[c]{\(p_T>20\,\mathrm{GeV}\)\\\(|\eta|<2.1\)}
        & 60--120
        & \(|y|<2.1\)
        & 4 \\
        \hline
        \makecell[c]{CMS 8 TeV~\cite{CMS:2016mwa}}
        & p-p
        & \(\sigma^{-1}\,\mathrm d\sigma/\mathrm dq_T\)
        & 8000
        & \makecell[c]{\(p_T>20\,\mathrm{GeV}\)\\\(|\eta|<2.1\)}
        & 60--120
        & \(|y|<2.1\)
        & 4 \\
        \hline
        \multirow[c]{8}{*}{\makecell[c]{CMS 13 TeV~\cite{CMS:2019raw}}}
        & \multirow[c]{8}{*}{p-p}
        & \multirow[c]{8}{*}{\(\mathrm d\sigma/\mathrm dq_T\)}
        & \multirow[c]{8}{*}{13000}
        & \multirow[c]{5}{*}{\makecell[c]{\(p_T>25\,\mathrm{GeV}\)\\\(|\eta|<2.4\)}}
        & \multirow[c]{5}{*}{76--106}
        & \(|y|<0.4\) & 13 \\
        & & & & & & \(0.4<|y|<0.8\) & 13 \\
        & & & & & & \(0.8<|y|<1.2\) & 13 \\
        & & & & & & \(1.2<|y|<1.6\) & 13 \\
        & & & & & & \(1.6<|y|<2.4\) & 13 \\
        \cline{5-7}
        & & & & \multirow[c]{3}{*}{\makecell[c]{\(p_{T,1}>25\,\mathrm{GeV}\)\\\(p_{T,2}>20\,\mathrm{GeV}\)\\\(|\eta|<2.4\)}}
        & \multirow[c]{3}{*}{\makecell[c]{106--170\\170--350\\350--1000}}
        & \multirow[c]{3}{*}{\(|y|<2.4\)} & 8 \\
        & & & & & & & 10 \\
        & & & & & & & 11 \\
        \hline
        \makecell[c]{LHCb 7 TeV~\cite{LHCb:2015okr}}
        & p-p
        & \(\mathrm d\sigma/\mathrm dq_T\)
        & 7000
        & \makecell[c]{\(p_T>20\,\mathrm{GeV}\)\\\(2<\eta<4.5\)}
        & 60--120
        & \(2<y<4.5\)
        & 7 \\
        \hline
        \makecell[c]{LHCb 8 TeV~\cite{LHCb:2015mad}}
        & p-p
        & \(\mathrm d\sigma/\mathrm dq_T\)
        & 8000
        & \makecell[c]{\(p_T>20\,\mathrm{GeV}\)\\\(2<\eta<4.5\)}
        & 60--120
        & \(2<y<4.5\)
        & 7 \\
        \hline
        \makecell[c]{LHCb 13 TeV~\cite{LHCb:2016fbk}}
        & p-p
        & \(\mathrm d\sigma/\mathrm dq_T\)
        & 13000
        & \makecell[c]{\(p_T>20\,\mathrm{GeV}\)\\\(2<\eta<4.5\)}
        & 60--120
        & \(2<y<4.5\)
        & 7 \\
        \hline
        Total & & & & & & & 267 \\
        \hline
    \end{tabular}
    \caption{Summary of collider Drell--Yan datasets included in the fit
    after the kinematic cut.}
    \label{tab:data_selection_collider}
\end{table}

For fixed-target Drell--Yan measurements, the fit includes E288 and E605,
together with the high-mass E772 subset with \(11<Q<15~\mathrm{GeV}\). At
collider energies, we include RHIC \(Z\)-boson data from STAR, Tevatron
measurements from CDF and D0, and neutral-current Drell--Yan/\(Z\)-boson
datasets from ATLAS, CMS, and LHCb. The retained fixed-target and collider
datasets are summarized separately in
tables~\ref{tab:data_selection_fixed} and
\ref{tab:data_selection_collider}, where each dataset group is annotated with
the corresponding original experimental reference.

\begin{table}[t]
    \centering
    \footnotesize
    \setlength{\tabcolsep}{2pt}
    \renewcommand{\arraystretch}{1.10}
    \begin{tabular}{|c|c|c|c|c|c|c|}
        \hline
        Dataset & Process & Observable & \(\sqrt{s}\) [GeV] & \(Q\) [GeV] & \(y\) or \(x_F\) & \(N_{\rm pts}\) \\
        \hline
        \makecell[c]{E288 (200 GeV)~\cite{Ito:1980ev}}
        & p-Cu
        & \makecell[c]{\(E\,\mathrm d\sigma/\mathrm d^3 q\)}
        & 19.4
        & \makecell[c]{4--5\\5--6\\6--7\\7--8\\8--9}
        & \(0.1<y<0.7\)
        & \makecell[c]{3\\4\\6\\7\\7} \\
        \hline
        \makecell[c]{E288 (300 GeV)~\cite{Ito:1980ev}}
        & p-Cu
        & \makecell[c]{\(E\,\mathrm d\sigma/\mathrm d^3 q\)}
        & 23.7
        & \makecell[c]{4--5\\5--6\\6--7\\7--8\\8--9\\11--12}
        & \(-0.09<y<0.51\)
        & \makecell[c]{3\\4\\6\\7\\7\\9} \\
        \hline
        \makecell[c]{E288 (400 GeV)~\cite{Ito:1980ev}}
        & p-Cu
        & \makecell[c]{\(E\,\mathrm d\sigma/\mathrm d^3 q\)}
        & 27.4
        & \makecell[c]{5--6\\6--7\\7--8\\8--9\\11--12\\12--13\\13--14}
        & \(-0.27<y<0.33\)
        & \makecell[c]{4\\6\\7\\7\\10\\12\\11} \\
        \hline
        \makecell[c]{E605~\cite{Moreno:1990sf}}
        & p-Cu
        & \makecell[c]{\(E\,\mathrm d\sigma/\mathrm d^3 q(x_F)\)}
        & 38.8
        & \makecell[c]{7--8\\8--9\\10.5--11.5\\11.5--13.5\\13.5--18}
        & \(-0.1<x_F<0.2\)
        & \makecell[c]{7\\7\\10\\11\\13} \\
        \hline
        \makecell[c]{E772~\cite{E772:1994cpf}}
        & p-D
        & \makecell[c]{\(E\,\mathrm d\sigma/\mathrm d^3 q(x_F)\)}
        & 38.8
        & \makecell[c]{11--12\\12--13\\13--14\\14--15}
        & \(0.1<x_F<0.3\)
        & \makecell[c]{8\\9\\7\\6} \\
        \hline
        Total & & & & & & 198 \\
        \hline
    \end{tabular}
    \caption{Summary of fixed-target Drell--Yan datasets included in the fit after the kinematic cut.}
    \label{tab:data_selection_fixed}
\end{table}

Since the factorized description of sec.~\ref{sec:formalism} is intended for
the low-$q_T$ regime, we retain only bins inside the corresponding TMD domain.
The universal kinematic selection is imposed bin by bin through
\begin{equation}
\frac{q_T^{\max}}{Q^{\min}}<0.2,
\end{equation}
with \(q_T^{\max}\) taken from the upper edge of the transverse-momentum bin
and \(Q_{\min}\) from the lower edge of the dilepton-mass bin. This choice
ensures that the full retained bin remains in the low-$q_T$ region and
therefore implements the cut conservatively. The final fitted dataset contains
\(N_{\rm dat}=465\) points.

Tables~\ref{tab:data_selection_fixed} and
\ref{tab:data_selection_collider} list the hadronic process, measured
observable, center-of-mass energy, mass range, kinematic coverage, and number
of fitted points for the retained dataset groups. The treatment of normalized
cross sections and nuclear targets is described in sec.~\ref{sec:fit-setup}.

\section{Fit setup}
\label{sec:fit-setup}


\subsection{Predictions for the measured observables}

In sec.~\ref{sec:formalism}, we derived the factorized expression for the
Drell--Yan differential cross section 
\(\mathrm d\sigma/(\mathrm dq_T\,\mathrm dQ\,\mathrm dy)\). We now specify how
the observables reported by each experiment are obtained from this quantity.

For the E288 measurement, the reported observable is
\(E\,\mathrm d\sigma/\mathrm d^3 q\), with bins defined directly in rapidity.
For a given bin, the theoretical prediction is
\begin{equation}
\left(E\frac{\mathrm d\sigma}{\mathrm d^3 q}\right)^{\rm th}_{\rm bin}
=
\frac{1}{\Delta q_T\,\Delta y}
\int_{q_T^{\min}}^{q_T^{\max}}\!\mathrm dq_T
\int_{Q^{\min}}^{Q^{\max}}\!\mathrm dQ
\int_{y^{\min}_{\rm phy}(Q)}^{y^{\max}_{\rm phy}(Q)}\!\mathrm dy\,
\frac{1}{2\pi q_T}\,
\frac{\mathrm d\sigma}{\mathrm dq_T\,\mathrm dQ\,\mathrm dy}\,.
\label{eq:pred_Edsigmad3q}
\end{equation}
Here \(y^{\min}_{\rm phy}(Q)\) and \(y^{\max}_{\rm phy}(Q)\) denote the lower
and upper boundaries of the set
\begin{equation}
\left\{
y\in\left(y^{\min},y^{\max}\right)
\,\middle|\,
0<x_1(y,Q),\,x_2(y,Q)<1
\right\},
\end{equation}
so that the rapidity integration is restricted to the kinematically allowed
region within the experimental bin. The integrated cross section is divided by
\(\Delta q_T\,\Delta y\).

For the E605 and E772 measurements, the observable is
\(E\,\mathrm d\sigma/\mathrm d^3 q(x_F)\), where the bins are defined in the
Feynman variable \(x_F\),
\begin{equation}
x_F=x_1-x_2=\frac{2Q\sinh y}{\sqrt{s}}\,.
\label{eq:xf_definition}
\end{equation}
At fixed \(Q\), the \(x_F\) bin can therefore be mapped to rapidity through
\begin{equation}
y(Q)
=
\operatorname{arsinh}\!\left(\frac{\sqrt{s}\,x_F}{2Q}\right),
\qquad
\frac{\partial x_F}{\partial y}
=
\frac{2Q\cosh y}{\sqrt{s}}\,.
\label{eq:xf_to_y}
\end{equation}
The corresponding theoretical prediction is
\begin{equation}
\left(E\frac{\mathrm d\sigma}{\mathrm d^3 q}(x_F)\right)^{\rm th}_{\rm bin}
=
\frac{1}{\Delta q_T\,\Delta x_F}
\int_{q_T^{\min}}^{q_T^{\max}}\!\mathrm dq_T
\int_{Q^{\min}}^{Q^{\max}}\!\mathrm dQ
\int_{y^{\min}_{\rm phy}(Q)}^{y^{\max}_{\rm phy}(Q)}\!\mathrm dy\,
\frac{1}{2\pi q_T}\,
\frac{2Q\cosh y}{\sqrt{s}}\,
\frac{\mathrm d\sigma}{\mathrm dq_T\,\mathrm dQ\,\mathrm dy}\,.
\label{eq:pred_Edsigmad3q_xf}
\end{equation}

For a nuclear target with mass number \(A\) and atomic number \(Z\), we model
its TMDs as a proton--neutron mixture,
\begin{equation}
F_q^A
=
\frac{Z}{A}\,F_q^p
+
\frac{A-Z}{A}\,F_q^n,
\label{eq:isospin_rotation_target}
\end{equation}
where the neutron distributions are obtained from the proton ones by the usual
\(u\leftrightarrow d\) isospin rotation. In this work, we do not include
nuclear PDFs or any additional nuclear corrections. The same treatment of
nuclear targets was adopted in \cite{Moos:2023yfa}.

For collider measurements whose observable is
\(\mathrm d\sigma/\mathrm dq_T\), the theoretical prediction is computed as
\begin{equation}
\left(\frac{\mathrm d\sigma}{\mathrm dq_T}\right)^{\rm th}_{\rm bin}
=
\frac{1}{\Delta q_T}
\int_{q_T^{\min}}^{q_T^{\max}}\!\mathrm dq_T
\int_{Q^{\min}}^{Q^{\max}}\!\mathrm dQ
\int_{y^{\min}_{\rm phy}(Q)}^{y^{\max}_{\rm phy}(Q)}\!\mathrm dy\,
\frac{\mathrm d\sigma}{\mathrm dq_T\,\mathrm dQ\,\mathrm dy}\,.
\label{eq:pred_dsigmadqT_abs}
\end{equation}
In this case, only the \(q_T\)-bin width appears in the denominator. The
original 7 and 8 TeV LHC data were not divided by their corresponding
\(q_T\)-bin widths, so we divide those data by \(\Delta q_T\) in order to make
their definition consistent with the observable
\(\mathrm d\sigma/\mathrm dq_T\).

For a dataset \(f\) reporting the normalized observable
\((1/\sigma)\,\mathrm d\sigma/\mathrm dq_T\), we approximate the total cross
section entering the normalization by
\begin{equation}
\sigma_f^{\rm th}
=
\frac{\sum_{i\in f}\Delta q_{T,i}\,
\left(
\frac{\mathrm d\sigma}{\mathrm dq_T}
\right)^{\rm th}_i}
{\sum_{i\in f}\Delta q_{T,i}\,
\left(
\frac{1}{\sigma}\frac{\mathrm d\sigma}{\mathrm dq_T}
\right)^{\rm exp}_i}\,,
\label{eq:approx_total_xsec}
\end{equation}
where \(\Delta q_{T,i}=q_{T,i}^{\max}-q_{T,i}^{\min}\), and the sum runs over
all fitted data points in dataset \(f\). The same
treatment of normalized cross sections was used in \cite{Moos:2023yfa}. The normalized theoretical prediction
for each bin is then taken as
\begin{equation}
\left(
\frac{1}{\sigma}\frac{\mathrm d\sigma}{\mathrm dq_T}
\right)^{\rm th}_i
=
\frac{1}{\sigma_f^{\rm th}}
\left(
\frac{\mathrm d\sigma}{\mathrm dq_T}
\right)^{\rm th}_i\,.
\end{equation}

\subsection{\texorpdfstring{$\chi^2$}{chi2} definition}

For each fitted data set \(f\), we construct the experimental covariance matrix as
\begin{equation}
V_f^{\rm exp}
=
V_f^{\rm unc}
+
\sum_k v_{f,k}^{\rm add}(v_{f,k}^{\rm add})^T
+
\sum_\ell v_{f,\ell}^{\rm mult}(v_{f,\ell}^{\rm mult})^T .
\end{equation}
Here \(V_f^{\rm unc}\) is diagonal and collects all uncorrelated contributions
in quadrature, while \(v_{f,k}^{\rm add}\) and \(v_{f,\ell}^{\rm mult}\)
denote correlated additive and multiplicative sources, respectively. For quoted
relative uncertainties, the corresponding absolute shifts are obtained from the
measured central values \(d_{f,i}\) according to
\begin{equation}
\bigl(v_{f,\ell}^{\rm mult}\bigr)_i
=
\delta_{f,\ell,i}^{\rm mult}\, d_{f,i}\,.
\end{equation}

To construct the covariance matrices associated with collinear PDF
uncertainties, \(V_f^{\rm PDF}\), we first convert the
MSHT20aN\textsuperscript{3}LO Hessian error set into a replica set with
\(N_{\rm mem}=100\) members using the public \texttt{MCGEN} code
\cite{Hou:2016sho}. At a fixed reference parameter point \(\theta_{\rm ref}\),
we compute the predictions \(t_f^{(a)}(\theta_{\rm ref})\) for each data set
\(f\) and each PDF member \(a=1,\ldots,N_{\rm mem}\), and define the shift
vectors with respect to the mean prediction,
\begin{equation}
\Delta_f^{(a)}(\theta_{\rm ref})
=
t_f^{(a)}(\theta_{\rm ref}) - \langle t_f(\theta_{\rm ref}) \rangle .
\end{equation}
The propagated collinear PDF covariance matrix is then given by
\begin{equation}
V_f^{\rm PDF}
=
\frac{1}{N_{\rm mem}-1}
\sum_{a=1}^{N_{\rm mem}}
\Delta_f^{(a)}(\theta_{\rm ref})
\,
\Delta_f^{(a)}(\theta_{\rm ref})^T .
\label{eq:collinear_pdf_covariance}
\end{equation}
A similar treatment of collinear PDF uncertainties in TMD extractions was
adopted by the MAP collaboration, as discussed in \cite{Bacchetta:2022awv}. However, we do not further decompose the propagated collinear PDF
uncertainty into separate correlated and uncorrelated components; instead, we
retain the full covariance matrix generated by the replica ensemble.

Strictly speaking, \(V_f^{\rm PDF}\) inherits a dependence on the fit
parameters through the theory predictions. In the present analysis, however, it
is evaluated once at \(\theta_{\rm ref}\) and then kept fixed throughout the
minimization. In practice, this amounts to treating the propagated collinear
PDF uncertainty as an external theory uncertainty defined around a nominal
reference point. This keeps the covariance matrix independent of the fit parameters and allows the minimization to be performed with the standard least-squares objective at manageable numerical cost. Since our aim is to propagate the prior collinear PDF uncertainty,
rather than to refit the collinear PDFs simultaneously with the TMD
parameters~\cite{Barry:2025glq}, this fixed-point construction provides a controlled approximation
for the central fit.

The full covariance matrix for data set \(f\) is then taken to be
\begin{equation}
V_f = V_f^{\rm exp}+V_f^{\rm PDF}.
\end{equation}
The fit quality for each data set is quantified by the chi-square
\begin{equation}
\chi_f^2(\theta)
=
\bigl(\vec d_f-\vec t_f(\theta)\bigr)^T
V_f^{-1}
\bigl(\vec d_f-\vec t_f(\theta)\bigr),
\end{equation}
and the total chi-square minimized in the fit is
\begin{equation}
\chi^2(\theta)
=
\sum_f \chi_f^2(\theta).
\end{equation}

\subsection{AI-driven exploration of the nonperturbative parameterizations}
\label{subsec:agents}

Rather than fixing the nonperturbative parameterization \emph{a priori}, we treated the
nonperturbative sector as a constrained ansatz space and explored it with OpenAI Codex (GPT-5.4) \cite{Chen:2021codex,OpenAI:GPT54Codex}. The search concerned two ingredients: the nonperturbative Sudakov factor
\(S_{\rm NP}(x,b)\) entering the TMD PDF and the nonperturbative contribution
\(D_{\rm NP}(b)\) to the Collins--Soper kernel. Their admissible forms were restricted by
the requirements of TMD factorization. For \(S_{\rm NP}\), we imposed
\begin{equation}
S_{\rm NP}(x,0)=1,\qquad
S_{\rm NP}(x,b)=1-\mathcal O(b^2)\quad (b\to0),\qquad
\lim_{b\to\infty}S_{\rm NP}(x,b)=0.
\label{eq:agent_constraints_snp}
\end{equation}
For \(D_{\rm NP}\), we required
\begin{equation}
D_{\rm NP}(b)=\mathcal O(b^2)\quad (b\to0).
\label{eq:agent_constraints_dnp}
\end{equation}
These conditions preserve the perturbative small-\(b\) limit while allowing sufficient
flexibility at intermediate and large \(b\).

\begin{figure}[t]
    \centering
    \makebox[\textwidth][c]{\includegraphics[width=1.1\textwidth]{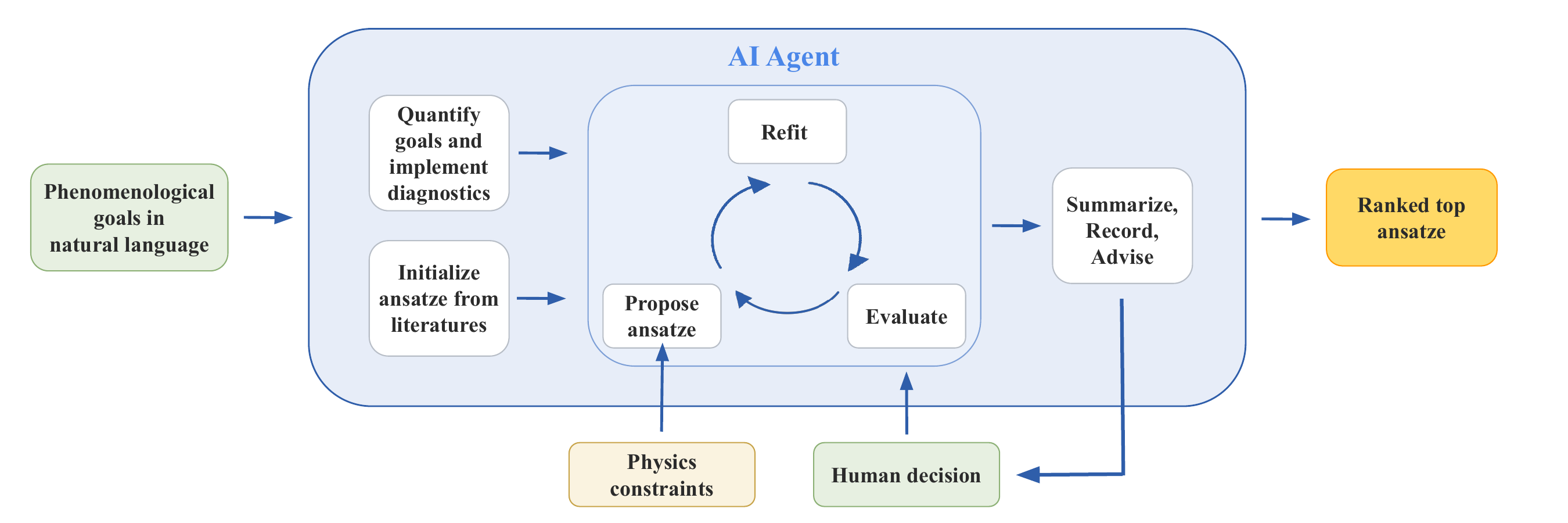}}
\caption{Schematic overview of the AI-agent-driven workflow for exploring nonperturbative ansatze. The blue box highlights the tasks carried out by the AI agent.}
\label{fig:flow}
\end{figure}

The exploration was initialized by human guidance expressed in natural language at the
level of phenomenological goals, rather than by explicit coding instructions. We asked the agent
to start from functional forms that can be found in existing TMD extractions and to improve the fit
with particular attention to collider datasets' normalization, the physical behavior of the
resulting TMD PDFs in $k_T$-space, and the avoidance of strongly correlated
parameterizations. The human role was to define the admissibility criteria and physics
priorities; the agent translated them into numerical diagnostics, constructed candidate
ansatze, and ran bounded multistart refits. 

From these instructions, the agent built a common diagnostic standard that drove an
iterative proposal--refit--evaluation loop. Global fit quality was evaluated through
\(\chi^2/N_{\rm pt}\), as usual. To target collider normalization more directly, the agent defined
a restricted set \(H\) consisting of the first three \(q_T\) bins of CMS 13 TeV
\((170<Q<350)\), CMS 13 TeV \((350<Q<1000)\), ATLAS 8 TeV \((|y|<0.4)\), ATLAS 8 TeV
\((0.4<|y|<0.8)\), and LHCb 8 TeV. For each bin \(i\in H\), it introduced the aggregate absolute normalization
deviation and aggregate shortfall,
\begin{equation}
\Delta_{\rm abs}^{H}
=
\sum_{i\in H}|r_i-1|,
\qquad
\Delta_{\rm short}^{H}
=
\sum_{i\in H}\max(0,1-r_i),
\label{eq:agent_delta_short}
\end{equation}
together with \(\Delta_{\rm short}^{H}\) isolates systematic underprediction. Here, \(r_i=t_i/d_i\) represent the theory--data ratio. The agent also designed finite-difference Hessian
analysis of parameter significance and correlations, and coded TMD plotters to verify that the
\(k_T\)-space TMD PDFs remained positive and that the \((x,b)\) and \((x,k_T)\) surfaces
stayed smooth.

The ansatz exploration was organized around \emph{branches}, namely families of
parameterizations that shared the same overall structure and differed only in a limited terms. This branching picture was important because
it allowed the search to be incremental and interpretable, instead of jumping between
unrelated ansatze. In the early stage of the exploration, for example, one representative branch generated by the agent is a 
Gaussian-like family for \(D_{\rm NP}\):
\begin{equation}
D_{\rm NP}^{\rm Gauss}(b)
=
\frac{g_2^2}{4}\,b^2,
\label{eq:dnp_branch_gauss_base}
\end{equation}
with nearby variants obtained by introducing either a saturating profile
\(b_{\rm sat}(b)\) or a power-regulated profile \(b_{\rm pow}(b)\),
\begin{equation}
\begin{aligned}
D_{\rm NP}^{\rm Gauss,1}(b)
&=
\frac{g_2^2}{4}\,\bigl[b_{\rm sat}(b)\bigr]^2,
\qquad
D_{\rm NP}^{\rm Gauss,2}(b)
=
\frac{g_2^2}{4}\,\bigl[b_{\rm pow}(b)\bigr]^2.
\end{aligned}
\label{eq:dnp_branch_gauss_vars}
\end{equation}
Here
\(b_{\rm sat}(b)=b/\sqrt{1+(b/B_{\rm NP})^2}\)
and
\(b_{\rm pow}(b)=b[1+(b/B_{\rm NP})^2]^{(p-1)/2}\).
Another representative branch was ART-like,
\begin{equation}
D_{\rm NP}^{\rm ART}(b)
=
b\,b_{\rm sat}(b)
\left[
c_0+c_1L_b
\right],
\qquad
L_b=\ln\!\left(\frac{b_{\rm sat}(b)}{B_{\rm NP}}\right),
\label{eq:dnp_branch_art_base}
\end{equation}
with nearby variants obtained by extending or simplifying the logarithmic structure,
\begin{equation}
\begin{aligned}
D_{\rm NP}^{\rm ART,1}(b)
&=
b\,b_{\rm sat}(b)
\left[
c_0+c_1L_b+c_2L_b^2
\right],
\qquad
D_{\rm NP}^{\rm ART,2}(b)
=
b\,b_{\rm sat}(b)\,c_0 .
\end{aligned}
\label{eq:dnp_branch_art_vars}
\end{equation}
Likewise, within a given branch for \(S_{\rm NP}(x,b)\), the search proceeded through nearby
families that shared the same broad shape and differed only in a few basis choices or
deformation terms. One early branch, for example, used a broad backbone built from
combinations of \((1-x)\), \(x\), and \(\ln x\) for $\mathcal{S}(x)$, optionally supplemented by a smooth
deformation in \(\ln x\), such as $[\ln(x+x_0)-\ln(1+x_0)]
$. Within that neighborhood, the agent tested softened logarithms
and alternative basis replacements such as \(\sqrt{x}\) or \(\ln(1-x)\).

The diagnostics were used both within and across branches. Within a branch, they
identified which ingredients materially improved the fit and which mainly introduced
instability or redundant flexibility; across branches, they allowed us to reject families
whose apparent gains came at the cost of poor collider normalization, awkward TMD
behavior, or unfavorable parameter correlations. A representative example was the
introduction of a smooth Gaussian bump in \(\ln x\), which added shape flexibility to improve description for collider datasets
without spoiling physics requirements. 
The workflow required only limited human intervention. After each batch of tests--that is, several proposal--refit--evaluate cycles--the agent summarized the results and suggested a small number of promising directions for the next stage of exploration, while a human supervisor selected which direction to pursue. At the end of the search, the agent reviewed the explored ansatze and produced a prioritized list according to their overall performance under the evaluation criteria. The final parameterization adopted in this work is the endpoint of this iterative search, which examined more than a hundred candidate forms.

To summarize, within this iterative proposal--refit--evaluation loop, the agent handled the generation and implementation of candidate parameterizations, together with their evaluation, comparison, and bookkeeping under a common protocol. This was particularly important because it's difficult for a human alone to reliably retain and compare the full set of outcomes from a large scale exploration on the order of hundreds of candidate forms judged against multiple metrics. This workflow made a broader and more systematic exploration feasible and greatly accelerated the search, reducing what previously required weeks, and sometimes months, of manual iteration to a matter of days. The explicit forms retained in the final fit are given in Sec.~\ref{subsec:phenomenological-realization}.

\subsection{Replica analysis}

As a non-Bayesian control study, we estimate the uncertainties of the fitted
parameters and derived predictions with the Monte Carlo replica method
\cite{Ball:2009qv}. For each data set \(f\), pseudodata
replicas are generated by drawing Gaussian fluctuations according to the
experimental covariance matrix,
\begin{equation}
\vec d_f^{\,(r)}
=
\vec d_f + \delta \vec d_f^{\,(r)},
\qquad
\delta \vec d_f^{\,(r)}
\sim
\mathcal N\!\left(0,\,V_f^{\rm exp}\right),
\end{equation}
with \(r=1,\ldots,N_{\rm refit}\). In the present analysis we take
\(N_{\rm refit}=100\).

The effect of collinear PDF uncertainty is included by assigning to each
replica refit a collinear PDF member \(a_r\). In principle, this would require
recomputing the full theory prediction with that PDF member at every step of
the minimization. In practice, we approximate this dependence by a fixed
observable-space shift evaluated at the reference point \(\theta_{\rm ref}\),
\begin{equation}
\vec t_f^{\,(r)}(\theta)
\approx
\vec t_f(\theta)
+
\Delta_f^{(a_r)}(\theta_{\rm ref}),
\qquad
\Delta_f^{(a)}(\theta_{\rm ref})
=
\vec t_f^{\,(a)}(\theta_{\rm ref})
-
\langle\vec t_f^{\,}(\theta_{\rm ref})\rangle .
\end{equation}
Thus, for each replica refit, the central theory prediction is supplemented by
the precomputed shift associated with the selected collinear PDF member, while
the normalization prescription and kinematic treatment are kept identical to
those of the central analysis. This approximation captures the leading effect
of varying the collinear PDF input while keeping the replica study
computationally tractable. The replica parameters are obtained by minimizing
\begin{equation}
\chi_r^2(\theta)
=
\sum_f
\bigl(\vec d_f^{\,(r)}-\vec t_f^{\,(r)}(\theta)\bigr)^T
V_f^{-1}
\bigl(\vec d_f^{\,(r)}-\vec t_f^{\,(r)}(\theta)\bigr).
\end{equation}
In sec.~\ref{sec:results}, we present the central \(68\%\) intervals of the parameter and prediction distributions obtained from the replica analysis.

\section{Bayesian inference}
\label{sec:bayesian}

After presenting the replica analysis as a non-Bayesian control study, we turn to the Bayesian inference of the nonperturbative parameters defined in eq.~\eqref{eq:NP_parameters}. 
Whereas the replica method probes an ensemble of pseudodata refits, the
Bayesian analysis characterizes the posterior distribution implied by the
observed dataset and a specified prior.

Let
$\theta$ denote the full set of TMD fit parameters and let $\vec d$ denote the
experimental data. The posterior probability density for $\theta$ is given by
Bayes' theorem,
\begin{equation}
p(\theta \mid \vec d)
=
\frac{\mathcal L(\vec d \mid \theta)\,\pi(\theta)}
{\mathcal Z(\vec d)}.
\label{eq:bayes_theorem}
\end{equation}
Here $\pi(\theta)$ is the prior distribution, $\mathcal L(\vec d \mid \theta)$
is the likelihood, and $\mathcal Z(\vec d)$ is the Bayesian evidence.

The prior $\pi(\theta)$ represents our initial probabilistic description of the
TMD parameter space before conditioning on the fitted Drell--Yan data. In the
present analysis, the prior serves as the starting distribution over the
nonperturbative TMD parameters, while the likelihood updates this distribution
using the information contained in the measurements. The posterior
$p(\theta\mid \vec d)$ then gives the final distribution for
the TMD parameters after combining prior information with experimental
constraints.

For a fixed choice of collinear PDF input, the
natural Bayesian likelihood is built from the same Gaussian $\chi^2$ function
introduced in sec.~\ref{sec:fit-setup},
\begin{equation}
\mathcal L(\vec d \mid \theta,\phi)
\propto
\exp\!\left[-\frac{1}{2}\chi^2(\theta,\phi)\right],
\label{eq:like_with_phi}
\end{equation}
where $\phi$ denotes the collinear PDF degrees of freedom. Thus, at fixed
collinear PDF input, the Bayesian analysis is based on the same theory--data
mismatch measure as the replica analysis of sec.~\ref{sec:fit-setup}. The key
difference is that, in the Bayesian framework, the collinear PDF uncertainty is
incorporated directly at the likelihood level by marginalizing over $\phi$. The corresponding marginalized likelihood is
\begin{equation}
\mathcal L_{\rm marg}(\vec d \mid \theta)
=
\int d\phi\,
\mathcal L(\vec d \mid \theta,\phi)\,\pi(\phi),
\label{eq:phi_marginalization}
\end{equation}
where $\pi(\phi)$ denotes the prior distribution for the collinear PDF
parameters. In practice, we approximate this marginalization using the
replica representation of the collinear PDF uncertainty. If the prior over
$\phi$ is represented by a discrete ensemble of $N_{\rm rep}$ collinear PDF
replicas with equal weight, then eq.~\eqref{eq:phi_marginalization} becomes
\begin{equation}
\mathcal L_{\rm marg}(\vec d \mid \theta)
\approx
\frac{1}{N_{\rm rep}}
\sum_{a=1}^{N_{\rm rep}}
\mathcal L^{(a)}(\vec d \mid \theta)
=
\frac{1}{N_{\rm rep}}
\sum_{a=1}^{N_{\rm rep}}
\exp\!\left[-\frac{1}{2}\chi_a^2(\theta)\right],
\label{eq:replica_marginalized_like}
\end{equation}
where $\chi_a^2(\theta)$ is the chi-square obtained using the $a$th collinear
PDF replica in the theory prediction. Equivalently, the marginalized
log-likelihood may be written as
\begin{equation}
\log \mathcal L_{\rm marg}(\vec d \mid \theta)
=
\log\!\left[
\frac{1}{N_{\rm rep}}
\sum_{a=1}^{N_{\rm rep}}
\exp\!\left(-\frac{1}{2}\chi_a^2(\theta)\right)
\right].
\label{eq:loglike_marginalized}
\end{equation}
In this way, the uncertainty from the collinear PDFs is propagated directly
into the Bayesian posterior.

The remaining ingredient in eq.~\eqref{eq:bayes_theorem} is the evidence,
defined by
\begin{equation}
\mathcal Z(\vec d)
=
\int d\theta\,
\mathcal L_{\rm marg}(\vec d \mid \theta)\,\pi(\theta).
\label{eq:evidence_def}
\end{equation}
which normalizes the posterior. 
eq.~\eqref{eq:evidence_def} makes clear that Bayesian inference requires the
exploration of a multi-dimensional parameter space.
Since each point in parameter space requires a likelihood evaluation, and each
likelihood evaluation itself involves the marginalization over collinear PDF
replicas, the full Bayesian analysis is computationally demanding. Furthermore, the standard Markov Chain Monte Carlo sampling algorithm tends to take order millions of likelihood evaluations or more, depending on the fit problem. To combat this difficulty, we leverage both a Multilayer Perceptron (MLP) emulator to accelerate prediction computations, as well as an affine-invariant ensemble MCMC sampler \texttt{emcee}~\cite{Foreman-Mackey:2012any, Goodman:2010affine} to reduce the required likelihood samplings.

\subsection{AI-assisted multilayer perceptron (MLP) emulator}
\label{subsec:ml_surrogate}

Even with an efficient MCMC sampler, the Bayesian inference remains expensive.
Each posterior evaluation requires the full theory prediction over the fitted
dataset together with the PDF marginalization of
eq.~\eqref{eq:replica_marginalized_like}. To reduce the cost of each likelihood call, we therefore employ a neural-network emulator as a surrogate for the full TMD theory calculation.

Rather than emulating the raw prediction vector $\vec t(\theta)$, we emulate
the corresponding whitened residual vector. Let $V$ denote the full covariance
matrix entering the fit and let
\begin{equation}
V = L L^T
\label{eq:cholesky_cov}
\end{equation}
be its Cholesky decomposition, with $L$ lower triangular. We define
\begin{equation}
\vec z(\theta)
=
L^{-1}\bigl(\vec t(\theta)-\vec d\bigr).
\label{eq:whitened_residual_def}
\end{equation}
In this basis the chi-square becomes
\begin{equation}
\chi^2(\theta)=\vec z(\theta)^T\vec z(\theta),
\label{eq:chi2_whitened}
\end{equation}
so the likelihood can be evaluated directly from the emulator output without
reconstructing the full covariance-weighted quadratic form at each step.

The PDF marginalization also takes a simple form in the whitened basis. If
$\vec z_{\rm c}(\theta)$ denotes the whitened residual for the central PDF
choice and $\Delta \vec z^{\,(a)}$ the precomputed whitened shift of the
$a$th PDF replica, then
\begin{equation}
\vec z^{\,(a)}(\theta)
=
\vec z_{\rm c}(\theta)+\Delta \vec z^{\,(a)},
\qquad
\Delta \vec z^{\,(a)}=L^{-1}\Delta \vec t^{(a)},
\label{eq:whitened_pdf_shift}
\end{equation}
and the implemented marginalized likelihood reads
\begin{equation}
\log\mathcal L_{\rm marg}(\vec d\mid \theta)
=
\log\!\left[
\frac{1}{N_{\rm rep}}
\sum_{a=1}^{N_{\rm rep}}
\exp\!\left(
-\frac{1}{2}\,
\bigl\|\vec z_{\rm c}(\theta)+\Delta \vec z^{\,(a)}\bigr\|^2
\right)
\right].
\label{eq:implemented_loglike}
\end{equation}
This is the quantity evaluated throughout the MCMC run.

Let $\theta=(\theta_1,\dots,\theta_9)$ denote the set of free nonperturbative
fit parameters in the 9-dimensional parameterization. For both surrogate
training and posterior sampling, we work in normalized coordinates,
\begin{equation}
x_i
=
\frac{\theta_i-\theta_i^{\rm min}}
{\theta_i^{\rm max}-\theta_i^{\rm min}},
\qquad
0\le x_i \le 1,
\label{eq:normalized_coords}
\end{equation}
so that the relevant parameter space is the unit hypercube. Below, when a
theory quantity is evaluated at a normalized point, the corresponding physical
parameter point $\theta(x)$ is understood. The emulator training set is
constructed in this normalized parameter space. Each training sample is
therefore one 9-dimensional normalized parameter vector $x$, and its target is
the $N_{\rm data}=465$ component whitened residual vector $\vec z(\theta(x))$
evaluated with the exact theory code. To organize coverage within this
normalized space, we define the distance from a central point $x_{\rm c}$ by
\begin{equation}
r(x)=\|x-x_{\rm c}\|_{\infty}.
\label{eq:radius_def}
\end{equation}
A shell-aware Latin-hypercube design \cite{McKay:1979lhs} is first used to populate a semiglobal
region around the central fit.

To reduce the dimensionality of the target space, we compress the whitened
residual vectors with a fixed PCA basis \cite{Jolliffe:2016pca}. Let $\vec c(x)\in\mathbb R^{96}$ denote the retained
latent coefficients associated with the normalized input $x$. The reconstructed
prediction then takes the form
\begin{equation}
\vec z(x)
=
\vec \mu_z + U_{\rm PCA}\,\vec c(x),
\end{equation}
where $\vec \mu_z$ is the PCA mean vector and $U_{\rm PCA}$ is the matrix of the
retained principal components. In the production model we retain 96 latent
components, which preserve more than $99.99\%$ of the variance of the whitened
training targets. The network is therefore trained to learn the latent map
$x\mapsto \vec c(x)$ rather than the full 465-dimensional output directly.

The model improvement procedure is organized with a controller--executor--reviewer
workflow implemented using OpenAI Codex (GPT-5.4). Before each model improvement round, the current training pool is rescored using the deployed surrogate in order to construct an adaptive anchor set. At a fixed parameter point, let
$z_i^{\rm emu}$ and $z_i^{\rm truth}$ denote the $i$th components of the
emulator-predicted and truth-level whitened residual vectors, respectively. For
each training point, we evaluate both the whitened-space emulator root-mean-squared
error (RMSE) and the
induced shift in the mean whitened chi-square,
\begin{equation}
{\rm RMSE}_z
=
\left[
\frac{1}{N_{\rm data}}
\sum_{i=1}^{N_{\rm data}}
\left(z_i^{\rm emu}-z_i^{\rm truth}\right)^2
\right]^{1/2},
\end{equation}
and
\begin{equation}
\Delta\!\left(\chi^2/N_{\rm data}\right)
=
\left|
\frac{1}{N_{\rm data}}\sum_{i=1}^{N_{\rm data}}(z_i^{\rm emu})^2
-
\frac{1}{N_{\rm data}}\sum_{i=1}^{N_{\rm data}}(z_i^{\rm truth})^2
\right|.
\end{equation}
These diagnostics are normalized shell-by-shell in the distance variable $r(x)$
of eq.~\eqref{eq:radius_def} and combined into a single hardness score, which
is further upweighted near the largest-radius shells of the semiglobal region,
especially the boundary shells near its outer edge. The resulting high-score
points define the adaptive anchor set, namely, the subset of difficult training
points used to seed the next round of local data generation.

These anchors are then used to bias the next round of truth-level data
generation. Concretely, the new samples are drawn with a shell-wise
Latin-hypercube strategy \cite{McKay:1979lhs} that combines a global component, which preserves broad
coverage of each shell, with a local anchor-centered component, which places
additional samples in the neighborhoods of the highest-scoring anchors. In this
way, the controller uses the measured failure modes of the current surrogate to
steer the next set of appended data toward the regions where the model is least
reliable. The executor carries out the corresponding append and retraining
stages without modifying the underlying code, and the reviewer checks the
retrained model against a fixed set of validation criteria. If those quality
conditions are not met, the reviewer requests an additional round of anchor
construction, data generation, and retraining.

\begin{figure}[t]
\centering
\makebox[\textwidth][c]{\includegraphics[width=1.1\textwidth]{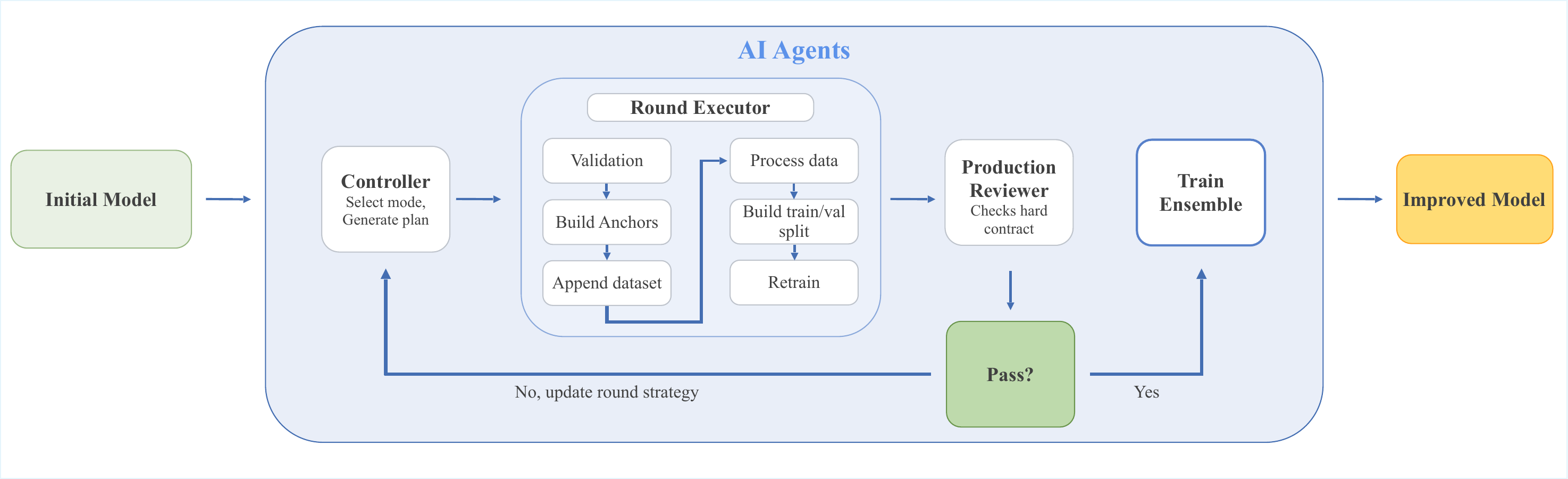}}
\caption{Schematic view of the controller--executor--reviewer workflow used for
adaptive anchor scoring, appended truth-level data generation, surrogate
retraining, and review against fixed validation criteria.
}
\label{fig:ML_flow}
\end{figure}

The AI assistance therefore enters through orchestration, adaptive data
selection, and model review rather than through unconstrained rewriting of the
theory calculation.

Model selection is not based only on the latent-space loss. In addition to the
validation latent mean-squared error (MSE), we monitor the reconstructed-space RMSE in $\vec z$, the
samplewise cosine similarity between truth and prediction, and the induced
$\chi^2/N_{\rm data}$ computed from the reconstructed whitened vectors.
Validation is also reported shell by shell in the same distance variable, so
that the same geometry used for adaptive data generation also enters the
model-quality assessment.

For posterior inference, we do not use a single deterministic surrogate output.
Instead, if $\widehat z_{k,i}(x)$ denotes the $i$th component of the prediction
of ensemble member $k$, we define
\begin{equation}
\overline z_i(x)
=
\frac{1}{N_{\rm ens}}
\sum_{k=1}^{N_{\rm ens}}\widehat z_{k,i}(x),
\label{eq:ensemble_mean}
\end{equation}
where $N_{\rm ens}$ denotes the number of ensemble members. The ensemble then gives a component-wise spread
\begin{equation}
\sigma_{z,i}(x)
=
\sqrt{
\operatorname{Var}_k\!\bigl[\widehat z_{k,i}(x)\bigr]
}.
\label{eq:ensemble_std}
\end{equation}
As in standard deep-ensemble approaches, the spread across ensemble members is
used here as a practical indicator of predictive uncertainty
\cite{Lakshminarayanan:2017deepensembles}. In the deployed production model, we use $N_{\rm ens}=8$ ensemble members, and
the central emulator prediction is given by $\overline{\vec z}(x)$.

Each ensemble member uses the same latent-space residual MLP architecture. The
production architecture uses width 384, four residual blocks, expansion factor
2, SiLU activations, LayerNorm, and dropout 0.02. The final model is trained on
a fixed 90\%/10\% train/validation split with the AdamW optimizer \cite{Loshchilov:2019adamw}, batch size
1024, initial learning rate $2\times 10^{-3}$, weight decay $10^{-4}$, and
cosine annealing to a minimum learning rate of $10^{-5}$ \cite{Loshchilov:2017sgdr}. Training is performed
in mixed precision, with a maximum budget of 300 epochs and early stopping
after 40 epochs without improvement in the validation metric.

To control surrogate reliability during sampling, we use a trust score based on
the ensemble spread,
\begin{equation}
u(x)
=
\left[
\frac{1}{N_{\rm data}}
\sum_{i=1}^{N_{\rm data}}
\sigma_{z,i}^2(x)
\right]^{1/2}.
\label{eq:gating_score}
\end{equation}
If $u(x)$ lies below a calibrated threshold, the marginalized likelihood is
evaluated from the emulator output. If $u(x)$ exceeds that threshold, the point
is reevaluated with the exact theory code and the likelihood is computed from
the truth-level result. In this way, the emulator accelerates the interior of
the posterior while retaining a controlled fallback in regions where the
ensemble itself signals reduced reliability.

\subsection{Posterior sampling with the affine-invariant ensemble sampler}
\label{subsec:emcee}

For posterior sampling, we work in the same normalized coordinates
$x_i$ defined in eq.~\eqref{eq:normalized_coords}. This is the same coordinate
system in which the emulator is trained and evaluated. In these coordinates,
the production runs employ an independent Beta prior on each component,
\begin{equation}
\pi(x)
=
\prod_{i=1}^{9}
\frac{x_i^{\alpha-1}(1-x_i)^{\beta-1}}{B(\alpha,\beta)},
\qquad
\alpha=\beta=2.
\label{eq:beta_prior}
\end{equation}
Uniform-prior runs are used only as a diagnostic of prior sensitivity.

The posterior is sampled with the affine-invariant ensemble package
\texttt{emcee}. This choice is well-suited to the present problem: the sampled
space is only moderately high-dimensional, but the posterior is correlated,
box-constrained, and still expensive to explore because each likelihood call
retains the PDF marginalization and, when needed, the truth-level fallback.

In normalized coordinates, the sampled posterior is
\begin{align}
\log p(x\mid \vec d)
=&
\log\mathcal L_{\rm marg}(\vec d\mid \theta(x))
\nonumber\\
&+
\sum_{i=1}^{9}
\left[
(\alpha-1)\log x_i + (\beta-1)\log(1-x_i) - \log B(\alpha,\beta)
\right]
+
{\rm const},
\label{eq:logposterior_normalized}
\end{align}
with $x_i\in[0,1]$ and $\alpha=\beta=2$ for the production runs. The ensemble
sampler avoids the need to hand-tune a fixed proposal covariance and is
therefore preferable to a basic random-walk Metropolis--Hastings chain in this
setting.

We evolve 64 walkers in the 9-dimensional parameter space. The walkers are initialized around
the central normalized point by
\begin{equation}
x_0^{(w)}
=
{\rm clip}\!\left[x_{\rm c}+\delta^{(w)},\,0,\,1\right],
\qquad
\delta^{(w)} \sim \mathcal N\!\left(0,\sigma_{\rm init}^2 I\right)\,,
\end{equation}
with $\sigma_{\rm init}=5\times 10^{-3}.$
The proposal kernel is taken to be a weighted mixture of differential-evolution
updates and snooker updates \cite{terBraak:2006demc,terBraak:2008snooker},
which improves exploration of correlated parameter space. The log-posterior is
evaluated in vectorized batches over the walker ensemble. The initial burn-in
phase is discarded in order to remove the transient dependence on walker
initialization. The production runs therefore discard the initial 1000 burn-in
steps. Thereafter the chains are advanced in chunks of 1000 steps and monitored
with three diagnostics: the integrated autocorrelation time $\tau_{\rm int}$,
the mean acceptance fraction, and the maximum split-$\hat{\mathcal R}$ across
all sampled parameters. Let $N_{\rm prod}$ denote the accumulated number of
post-burn production steps. Then the production loop is terminated once
$
N_{\rm prod}\ge 50\,\tau_{\rm int},
$
and
$
\max (\hat{\mathcal R}) < 1.01
$
\cite{Vehtari:2021rhat}.
This adaptive stopping criterion provides a practical compromise between cost
and posterior fidelity.

Posterior summaries are extracted from the flattened post-burn chain, with mild
thinning applied when useful for diagnostics and plotting. These retained
samples are then used to construct marginal parameter distributions,
parameter-correlation plots, and predictive uncertainty bands in sec.~\ref{sec:results}.

\section{Results}
\label{sec:results}

\subsection{Fit result}

\begin{table}[t]
\centering
\renewcommand{\arraystretch}{1.15}
\begin{tabular}{lccc}
\hline
Method & \(\lambda_1\) & \(\lambda_2\) & \(\lambda_3\) \\
\hline
Replica  & \(0.0164^{+0.031}_{-0.044}\) & \(1.03^{+0.14}_{-0.18}\) & \(-2.36^{+0.47}_{-0.40}\) \\
Bayesian & \(0.0116^{+0.034}_{-0.034}\) & \(1.16^{+0.18}_{-0.19}\) & \(-2.81^{+0.49}_{-0.47}\) \\
\hline
Method & \(x_0\) & \(\sigma_x\) & \(A_{\rm NP}\) \\
\hline
Replica  & \(0.00569^{+0.0019}_{-0.0016}\) & \(1.10^{+0.13}_{-0.11}\) & \(-0.473^{+0.15}_{-0.11}\) \\
Bayesian & \(0.00495^{+0.0020}_{-0.0015}\) & \(1.54^{+0.34}_{-0.34}\) & \(-0.475^{+0.14}_{-0.14}\) \\
\hline
Method & \(B_{\rm NP}\) & \(c_0\) & \(c_1\) \\
\hline
Replica  & \(1.53^{+0.14}_{-0.13}\) & \(0.0698^{+0.0056}_{-0.0061}\) & \(0.0282^{+0.0078}_{-0.0073}\) \\
Bayesian & \(1.81^{+0.14}_{-0.14}\) & \(0.0643^{+0.0033}_{-0.0034}\) & \(0.0251^{+0.0024}_{-0.0025}\) \\
\hline
\end{tabular}
\caption{Mean values of the nonperturbative parameters and their uncertainties.}
\label{tab:fit_parameters}
\end{table}

The extracted nonperturbative (NP) parameters are summarized in
table~\ref{tab:fit_parameters}. The quoted central values correspond to the
means of the replica and Bayesian distributions, while the uncertainties are
defined by the central 68\% intervals of the fitted parameter distributions.
Notably, the mean parameter values obtained from the two inference strategies
correspond to different local minima, though with comparable fit quality. The
best-fit point yields \(\chi^2/N=1.01\), while the replica and Bayesian mean
parameter sets give \(\chi^2/N=1.02\) and \(\chi^2/N=1.03\), respectively.
Several qualitative features are common to both analyses. For the parameters
governing the NP Sudakov shape function \(\mathcal{S}(x)\), \(\lambda_1\)
remains small and is compatible with zero within uncertainties, while
\(\lambda_2>0\) and \(\lambda_3<0\) in both cases, indicating that the data
favor a smooth \(x\)-dependence. The broad Gaussian deformation in \(\ln x\)
is centered at small \(x\sim 5\times 10^{-3}\), with amplitude
\(A_{\rm NP}\simeq -0.47\) and width greater than unity. For the NP
Collins--Soper kernel, both methods prefer positive coefficients \(c_0\) and
\(c_1\), together with \(B_{\rm NP}>b_0~\mathrm{GeV}^{-1}\). The largest
differences between the two methods occur in \(\sigma_x\) and \(B_{\rm NP}\):
the Bayesian fit favors both a broader deformation in \(\ln x\) and a larger
saturation scale, while the remaining parameters differ only moderately.

\begin{table}[t]
    \centering
    \small
    \renewcommand{\arraystretch}{1.1}
    \setlength{\tabcolsep}{4pt}
    \makebox[\textwidth][c]{%
    \begin{minipage}[t]{0.495\textwidth}
        \centering
        \vspace{0pt}
        \begin{tabular}{@{}lcc@{}}
            \toprule
            Dataset & $\chi^2_R/N$ & $\chi^2_B/N$ \\
            \midrule
            STAR & 1.81 & 1.93 \\
            \midrule
            CDF Run I & 0.57 & 0.58 \\
            CDF Run II & 0.82 & 0.81 \\
            D0 Run I & 0.57 & 0.54 \\
            D0 Run II & 0.92 & 0.92 \\
            D0 Run II ($\mu^+\mu^-$) & 0.53 & 0.43 \\
            \textbf{Tevatron total} & 0.69 & 0.67 \\
            \midrule
            ATLAS 7 TeV, $|y|<1$ & 0.64 & 0.60 \\
            ATLAS 7 TeV, $1<|y|<2$ & 3.53 & 3.56 \\
            ATLAS 7 TeV, $2<|y|<2.4$ & 1.29 & 1.52 \\
            ATLAS 8 TeV, $|y|<0.4$ & 2.63 & 2.74 \\
            ATLAS 8 TeV, $0.4<|y|<0.8$ & 1.03 & 0.88 \\
            ATLAS 8 TeV, $0.8<|y|<1.2$ & 0.55 & 0.70 \\
            ATLAS 8 TeV, $1.2<|y|<1.6$ & 1.93 & 1.56 \\
            ATLAS 8 TeV, $1.6<|y|<2.0$ & 0.94 & 0.84 \\
            ATLAS 8 TeV, $2.0<|y|<2.4$ & 1.01 & 0.77 \\
            ATLAS 8 TeV, $46<Q<66$ & 0.76 & 0.86 \\
            ATLAS 8 TeV, $116<Q<150$ & 0.61 & 0.57 \\
            \textbf{ATLAS total} & 1.35 & 1.32 \\
            \midrule
            CMS 7 TeV & 1.44 & 1.46 \\
            CMS 8 TeV & 0.68 & 0.74 \\
            CMS 13 TeV, $|y|<0.4$ & 1.39 & 1.37 \\
            CMS 13 TeV, $0.4<|y|<0.8$ & 0.85 & 0.83 \\
            CMS 13 TeV, $0.8<|y|<1.2$ & 0.45 & 0.46 \\
            CMS 13 TeV, $1.2<|y|<1.6$ & 0.21 & 0.22 \\
            CMS 13 TeV, $1.6<|y|<2.4$ & 0.23 & 0.23 \\
            CMS 13 TeV, $106<Q<170$ & 0.91 & 0.81 \\
            CMS 13 TeV, $170<Q<350$ & 1.14 & 1.10 \\
            CMS 13 TeV, $350<Q<1000$ & 1.55 & 1.54 \\
            \textbf{CMS total} & 0.83 & 0.82 \\
            \midrule
            LHCb 7 TeV & 0.92 & 0.92 \\
            LHCb 8 TeV & 0.47 & 0.49 \\
            LHCb 13 TeV & 0.73 & 0.75 \\
            \textbf{LHCb total} & 0.71 & 0.72 \\
            \midrule
            \textbf{Collider total} & 0.94 & 0.93 \\
            \bottomrule
        \end{tabular}
    \end{minipage}
    \hspace{0.01\textwidth}
    \begin{minipage}[t]{0.495\textwidth}
        \centering
        \vspace{0pt}
        \begin{tabular}{@{}lcc@{}}
            \toprule
            Dataset & $\chi^2_R/N$ & $\chi^2_B/N$ \\
            \midrule
            E288 (200 GeV), $4<Q<5$ & 0.40 & 0.55 \\
            E288 (200 GeV), $5<Q<6$ & 0.26 & 0.29 \\
            E288 (200 GeV), $6<Q<7$ & 0.50 & 0.60 \\
            E288 (200 GeV), $7<Q<8$ & 0.66 & 0.89 \\
            E288 (200 GeV), $8<Q<9$ & 0.50 & 0.61 \\
            E288 (300 GeV), $4<Q<5$ & 1.43 & 1.60 \\
            E288 (300 GeV), $5<Q<6$ & 0.27 & 0.19 \\
            E288 (300 GeV), $6<Q<7$ & 0.63 & 0.99 \\
            E288 (300 GeV), $7<Q<8$ & 0.22 & 0.24 \\
            E288 (300 GeV), $8<Q<9$ & 1.08 & 0.96 \\
            E288 (300 GeV), $11<Q<12$ & 0.41 & 0.43 \\
            E288 (400 GeV), $5<Q<6$ & 0.81 & 0.97 \\
            E288 (400 GeV), $6<Q<7$ & 1.78 & 2.42 \\
            E288 (400 GeV), $7<Q<8$ & 1.45 & 1.13 \\
            E288 (400 GeV), $8<Q<9$ & 1.65 & 1.61 \\
            E288 (400 GeV), $11<Q<12$ & 1.00 & 1.06 \\
            E288 (400 GeV), $12<Q<13$ & 1.09 & 1.03 \\
            E288 (400 GeV), $13<Q<14$ & 0.69 & 0.71 \\
            \midrule
            E605, $7<Q<8$ & 0.22 & 0.23 \\
            E605, $8<Q<9$ & 0.53 & 0.51 \\
            E605, $10.5<Q<11.5$ & 0.54 & 0.56 \\
            E605, $11.5<Q<13.5$ & 0.75 & 0.71 \\
            E605, $13.5<Q<18$ & 1.37 & 1.39 \\
            \midrule
            E772, $11<Q<12$ & 2.56 & 2.72 \\
            E772, $12<Q<13$ & 3.73 & 3.82 \\
            E772, $13<Q<14$ & 4.04 & 4.23 \\
            E772, $14<Q<15$ & 0.25 & 0.27 \\
            \midrule
            \textbf{Fixed-target total} & 1.12 & 1.18 \\
            \midrule            
            \textbf{All Data} & 1.02 & 1.03 \\
            \bottomrule
        \end{tabular}
    \end{minipage}}
    \caption{Summary of the fit quality by dataset for the replica and Bayesian methods. The quantities $\chi_R^2/N$ and $\chi_B^2/N$ denote the values of $\chi^2/N$ evaluated using the mean parameter sets obtained from the replica and Bayesian analyses, respectively.}
    \label{tab:chi2_breakdown}
\end{table}

The fit quality by dataset is summarized in
table~\ref{tab:chi2_breakdown}. The reported replica and Bayesian
\(\chi^2/N\) values are evaluated using the replica and Bayesian mean
parameter values, respectively. Both inference strategies provide an overall
description of the full dataset at the level of \(\chi^2/N\simeq 1\). On
average, the Tevatron, CMS, and LHCb subsets are described below the
level of \(\chi^2/N\sim 1\). The ATLAS 7 TeV bins, especially the
\(1<|y|<2\) subset, yield larger \(\chi^2/N\) values, which can be attributed
primarily to their sub-percent experimental uncertainties. The fixed-target
data are also described reasonably well overall, with \(\chi^2/N=1.12\) for
the replica fit and \(\chi^2/N=1.18\) for the Bayesian fit, although the E772
bins remain the most challenging part of the fitted dataset. The close
agreement between the replica and Bayesian \(\chi^2\) values throughout the
table indicates that the predictions are stable with respect to the choice of
inference framework.

\begin{figure}[t]
\centering
\makebox[\textwidth][c]{\includegraphics[width=1.05\textwidth]{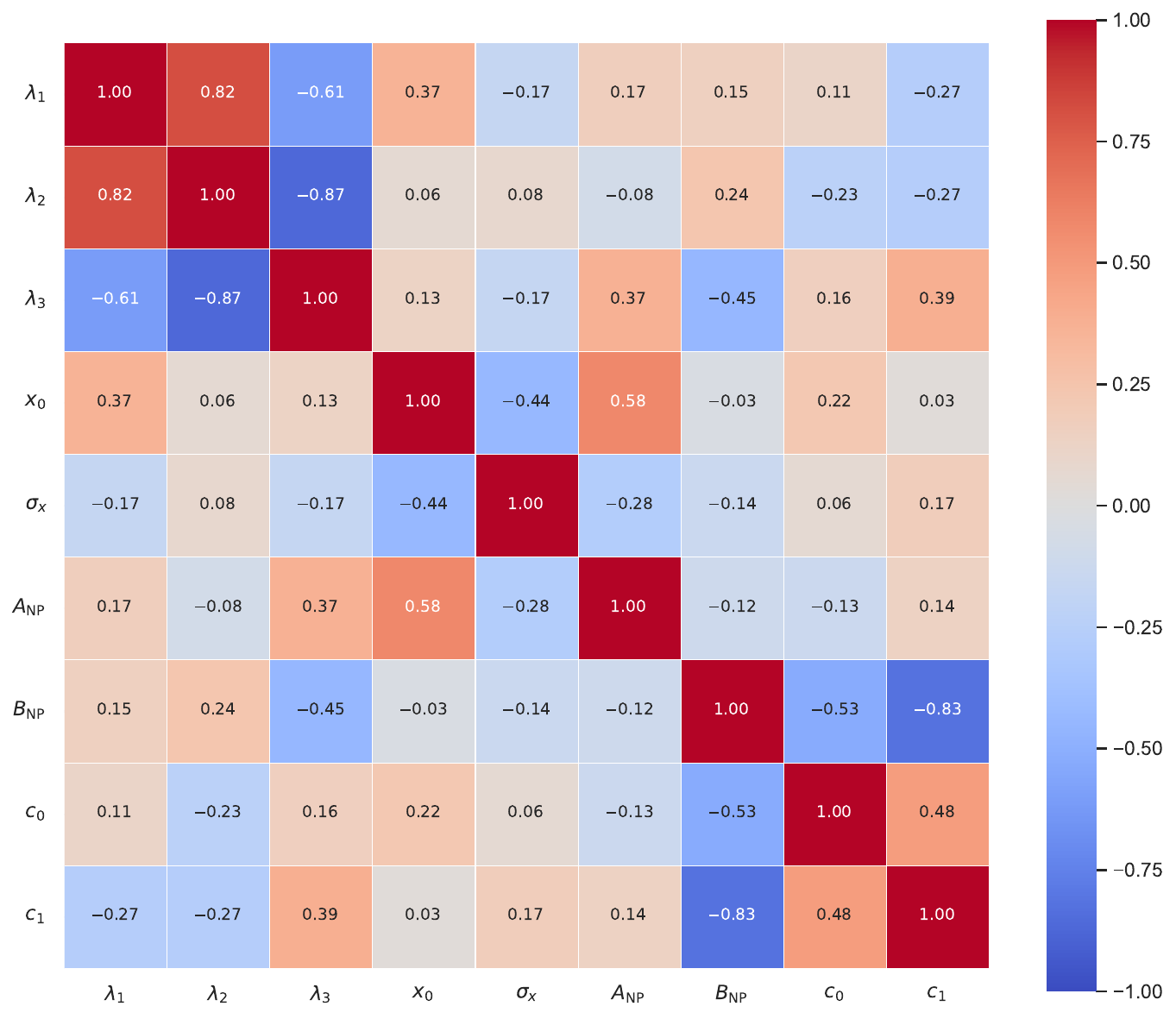}}
\caption{Correlation matrix for the replica distribution of the
nonperturbative parameters. The entries displayed in each cell are
the corresponding correlation coefficients.}
\label{fig:parameter_correlation_replica}
\end{figure}

The replica correlation matrix is shown in
fig.~\ref{fig:parameter_correlation_replica}. Overall, only three parameter pairs,
\(\lambda_1\)--\(\lambda_2\), \(\lambda_2\)--\(\lambda_3\), and
\(c_1\)--\(B_{\rm NP}\), exhibit strong correlations with magnitude above
0.7, while all other pairs are moderately or weakly correlated. The nine
fitted parameters can be grouped into three subsets: the shape subset
\(\{\lambda_1,\lambda_2,\lambda_3\}\), the Gaussian subset
\(\{A_{\rm NP},x_0,\sigma_x\}\), and the Collins--Soper subset
\(\{c_0,c_1,B_{\rm NP}\}\). As expected, correlations between different
subsets are weak, with the largest absolute correlation reaching only 0.45.
Within each subset, the matrix indicates that the data constrain particular
parameter combinations more strongly than individual coefficients. In
particular, the \(x\)-shape parameters \(\lambda_1\), \(\lambda_2\), and
\(\lambda_3\) are more strongly correlated than average, with a sizable
positive \(\lambda_1\)--\(\lambda_2\) correlation and a strong negative
\(\lambda_2\)--\(\lambda_3\) correlation. For the NP Collins--Soper kernel,
\(B_{\rm NP}\) is strongly anti-correlated with \(c_1\). The implications of
these fitted parameters for the extracted Collins--Soper kernel and TMD PDFs
are discussed in sec.~\ref{subsec:CS&TMDPDF}.

\begin{figure}[t]
\centering
\makebox[\textwidth][c]{\includegraphics[width=1.05\textwidth]{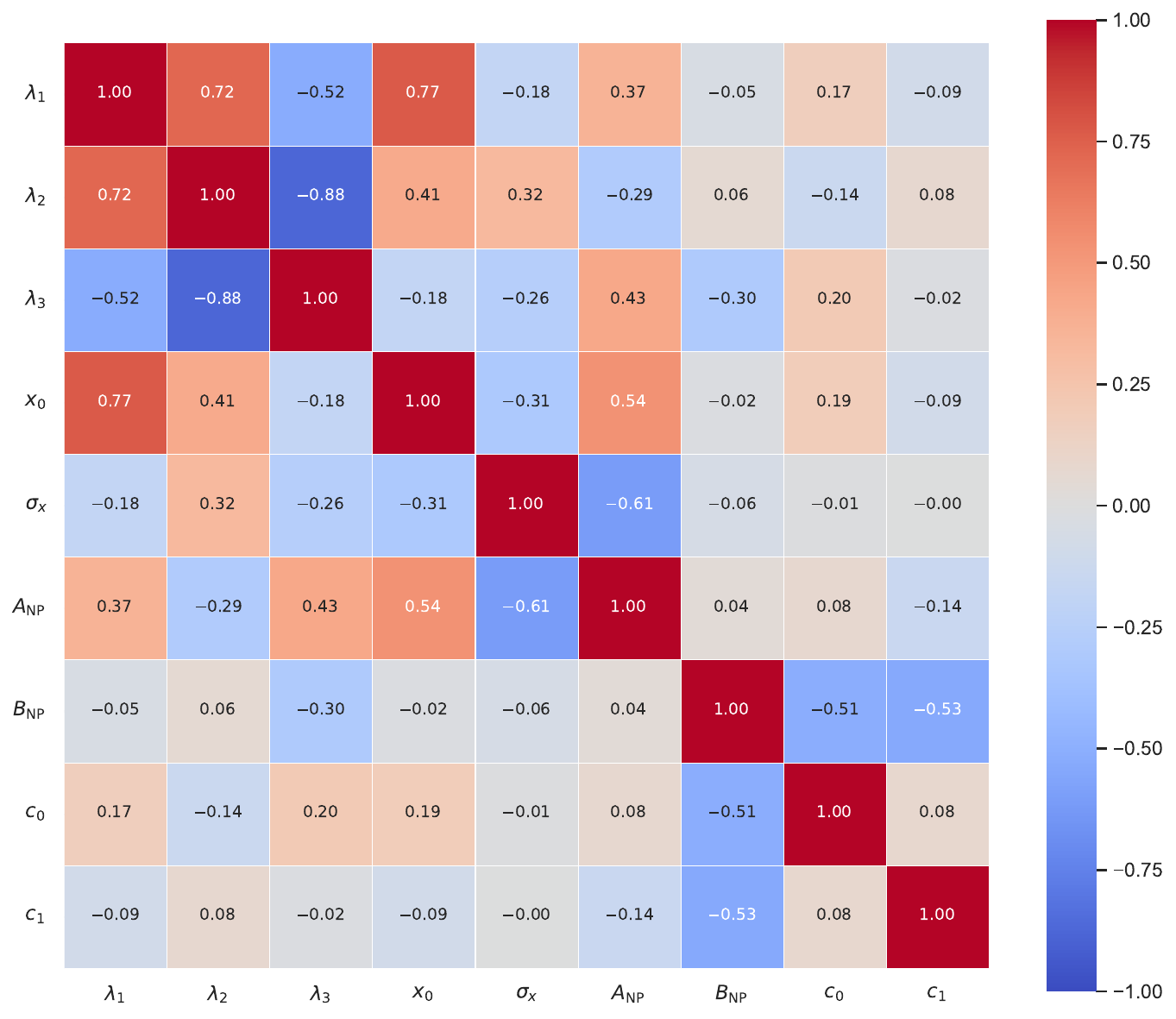}}
\caption{Correlation matrix for the Bayesian distribution of the
nonperturbative parameters. The entries displayed in each cell are
the corresponding correlation coefficients.}
\label{fig:parameter_correlation_bayesian}
\end{figure}

The Bayesian correlation matrix is shown in
fig.~\ref{fig:parameter_correlation_bayesian}. Its overall structure is
similar to that of the replica matrix. A complementary visualization of the full posterior distributions is provided by the corner plot in fig.~\ref{fig:cornerplot} in the appendix. In particular, the positive
\(\lambda_1\)--\(\lambda_2\) correlation and the strong negative
\(\lambda_2\)--\(\lambda_3\) correlation persist, indicating that both
inference strategies favor similar parameter combinations in the
\(x\)-shape sector. The main differences with respect to the replica case are twofold. First, the
Bayesian matrix shows a stronger interplay between the shape and Gaussian
subsets, most notably through the sizable positive
\(\lambda_1\)--\(x_0\) correlation. Second, the Collins--Soper sector is less
dominated by a single parameter pair: rather than a strong
\(c_1\)--\(B_{\rm NP}\) anti-correlation, the Bayesian analysis yields more
moderate anti-correlations of \(B_{\rm NP}\) with both \(c_0\) and \(c_1\).

\subsection{Extracted Collins--Soper kernel and TMD PDFs}
\label{subsec:CS&TMDPDF}

\begin{figure}[t]
\centering
\makebox[\textwidth][c]{\includegraphics[width=1.0\textwidth]{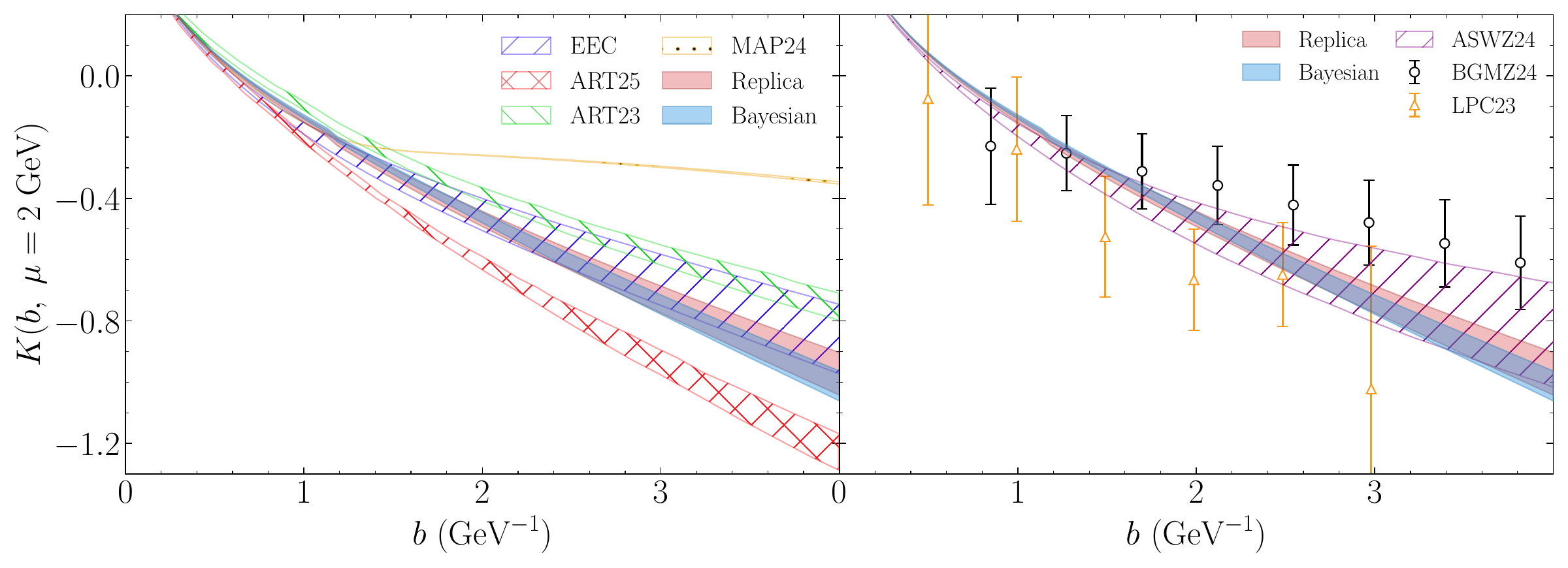}}
\caption{Comparison of the extracted Collins--Soper kernel with 
phenomenological and lattice determinations. Here, we plotted  $K(b,\mu) = -2D(b,\mu)$.}
\label{fig:cs_kernel_comparison}
\end{figure}

The extracted Collins--Soper kernel is shown in fig.~\ref{fig:cs_kernel_comparison} at \(\mu=2~\mathrm{GeV}\) for \(0<b<4~\mathrm{GeV}^{-1}\). Although the mean fitted parameters differ between the replica and Bayesian analyses, the resulting Collins--Soper kernels are compatible, with overlapping 68\% uncertainty bands over the full \(b\) range. In the left panel, we compare our extraction with other phenomenological determinations. The EEC result~\cite{Kang:2024dja} was obtained from fits to the back-to-back energy--energy correlator in \(e^+e^-\) annihilation. ART23~\cite{Moos:2023yfa} was extracted using Drell--Yan (DY) data only, while MAP24~\cite{Bacchetta:2024qre} and ART25~\cite{Moos:2025sal} included both semi-inclusive deep inelastic scattering (SIDIS) and DY data. Concerning the nonperturbative ansatz for the Collins--Soper kernel, the ART-series extractions used the same functional form as in the present work, whereas MAP24~\cite{Bacchetta:2024qre} adopted a simple \(b^2\) form and the EEC extraction used an ansatz proportional to \(b\,b^*\). In the right panel, we compare our result with lattice-QCD determinations. BGMZ24~\cite{Bollweg:2024zet} and LPC23~\cite{LatticePartonLPC:2023pdv} results were reported as discrete points, while ASWZ24~\cite{Avkhadiev:2024mgd} provided a continuum-extrapolated band.

At small \(b\), our extracted Collins--Soper kernel is consistent with the other phenomenological determinations, as expected in the region where the behavior is dominated by the perturbative rapidity anomalous dimension. In the moderate and large-\(b\) region, where the nonperturbative contribution becomes relevant, our result is closest to the EEC extraction, with overlapping 68\% bands. Relative to the ART series, our central band lies between the ART23~\cite{Moos:2023yfa} and ART25~\cite{Moos:2025sal} results, while MAP24~\cite{Bacchetta:2024qre} shows a milder falloff at moderate and large \(b\). On the lattice side, our result is in good agreement with ASWZ24, whose uncertainty band largely covers ours. It is also compatible with LPC23 within uncertainties, whereas BGMZ24 tends to favor a smaller kernel at large \(b\).

\begin{figure}[t]
\centering
\makebox[\textwidth][c]{\includegraphics[width=1.05\textwidth]{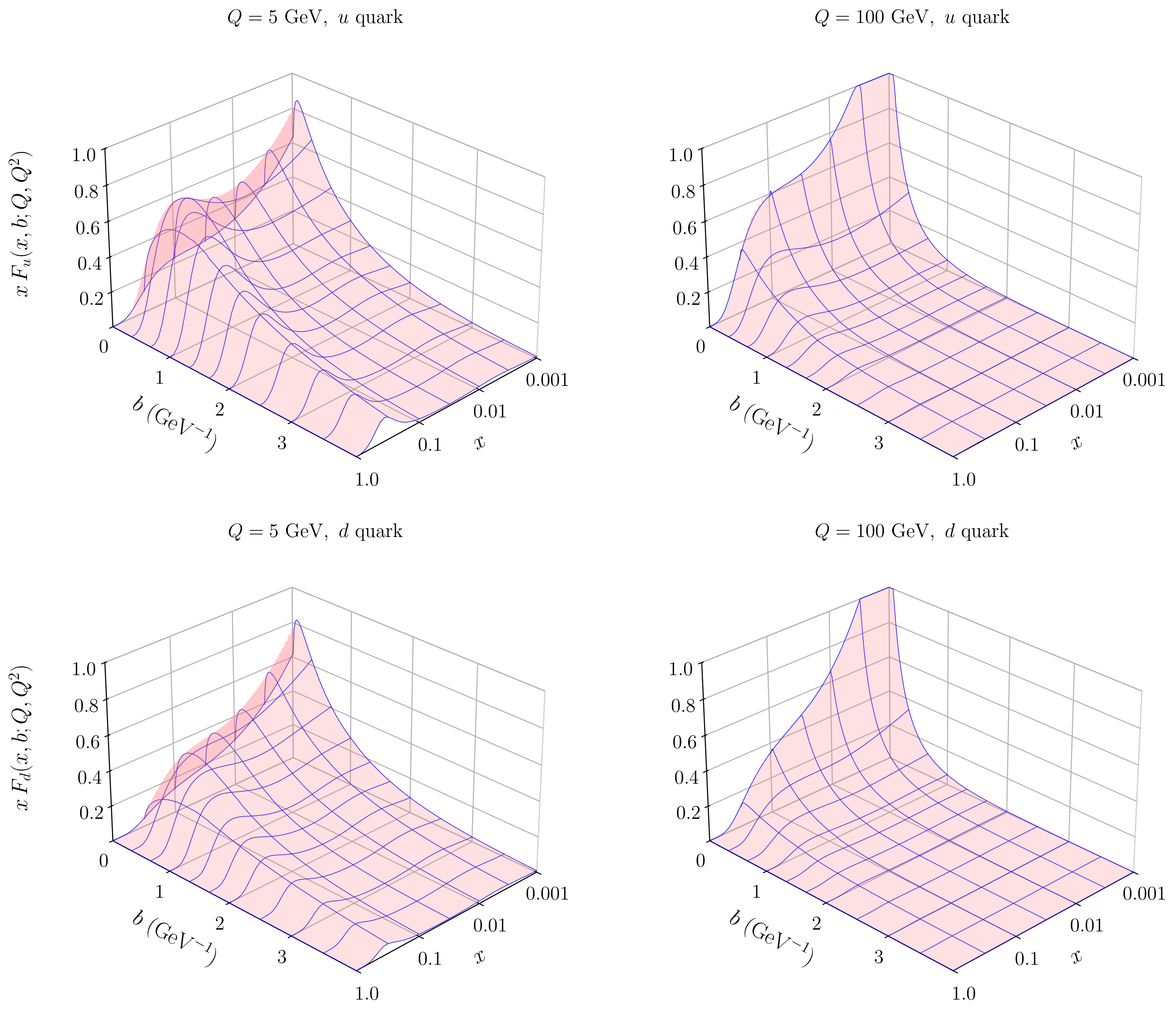}}
\caption{b-space TMD PDF \(x\,F_q(x,b;Q,Q^2)\)
for \(q=u,d\) at \(Q=5~\mathrm{GeV}\) and \(Q=100~\mathrm{GeV}\). Here, the plotted $b$ range is $0.01<b<4$.}
\label{fig:tmdpdf_u_b_surface}
\end{figure}

We present the \(b\)-space TMD PDFs in
fig.~\ref{fig:tmdpdf_u_b_surface} for the up and down quarks at
\(Q=5~\mathrm{GeV}\) and \(Q=100~\mathrm{GeV}\). The lower scale
\(Q=5~\mathrm{GeV}\) lies close to the lower end of the fixed-target \(Q\)
range included in the fit, while \(Q=100~\mathrm{GeV}\) is representative of
collider kinematics. In all panels, the extracted TMD PDFs are smooth
functions of both \(x\) and \(b\), with no localized spikes or oscillatory
behaviors. They exhibit the expected suppression at large \(b\), and this
suppression becomes more pronounced at higher \(Q\), consistent with the stronger suppression due to evolution over a larger Collins--Soper scale interval. At \(Q=5~\mathrm{GeV}\), the distributions are also suppressed near the lower
edge of the plotted \(b\) range, so that their maxima occur at intermediate
\(b\) rather than at \(b\to 0\). This feature originates from the
perturbative small-\(b\) region. In our implementation, the \(b^*\)
prescription regularizes the large-\(b\) behavior but does not introduce a
lower freezing scale at small \(b\). As a result, the canonical initial scale
\(\mu_i=b_0/b^*\) increases rapidly as \(b\) decreases. At the lower plotting
limit \(b=0.01~\mathrm{GeV}^{-1}\), one finds \(\mu_i\simeq 112~\mathrm{GeV}\).
The perturbative evolution from this scale down to \(Q=5~\mathrm{GeV}\)
therefore leads to suppression at very small \(b\). By
contrast, for \(Q=100~\mathrm{GeV}\) the corresponding evolution interval is
much shorter, and the small-\(b\) suppression is accordingly milder.
The \(x\)-dependence also evolves with the hard scale. At \(Q=5~\mathrm{GeV}\),
the TMD PDFs are larger at moderate \(x\), whereas after evolution
to \(Q=100~\mathrm{GeV}\) the ridge shifts toward smaller \(x\) for small $b$.

\begin{figure}[t]
    \centering
    \makebox[\textwidth][c]{\includegraphics[width=1.05\textwidth]{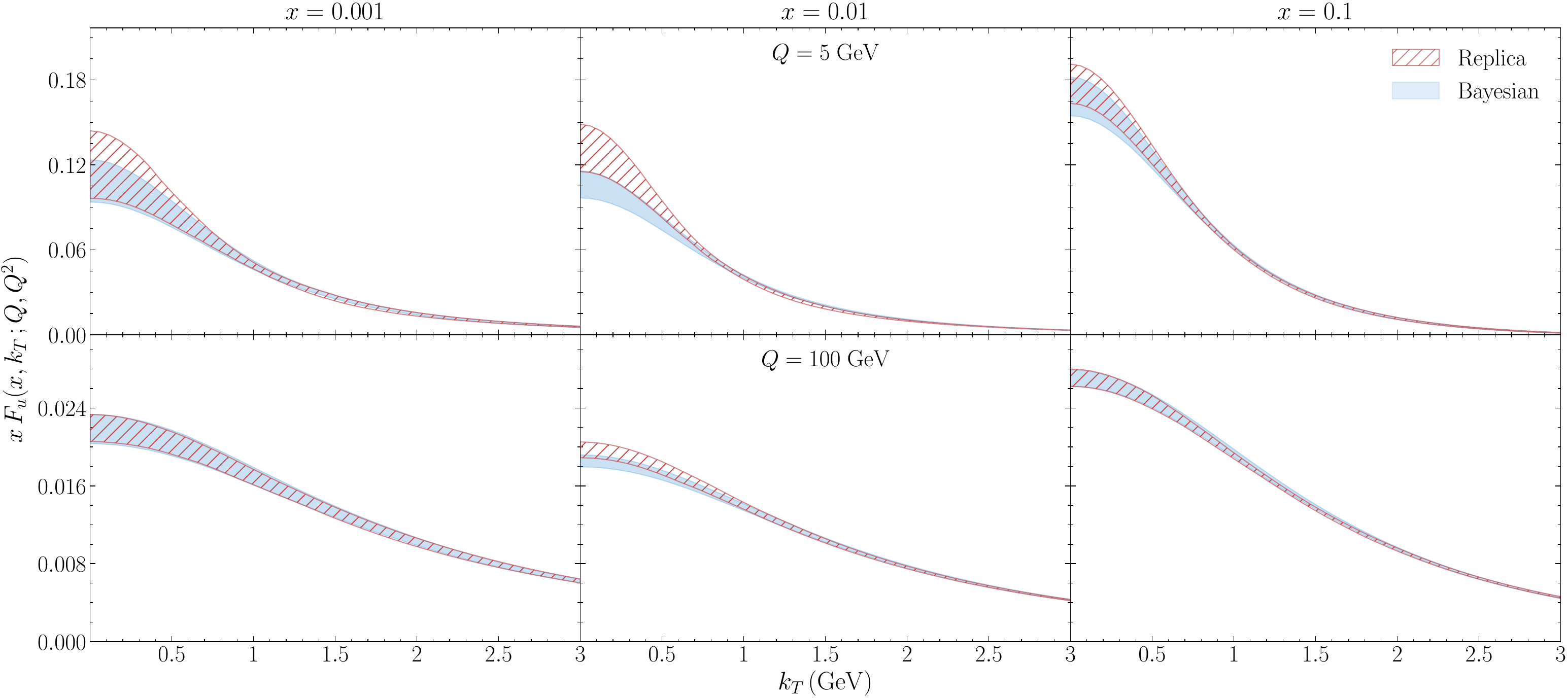}}
    \caption{Comparison of replica and Bayesian $68\%$ uncertainty bands for
    the momentum-space TMD PDF \(x\,F_u(x,k_T;Q,Q^2)\). The three columns
    correspond to \(x=0.001\), \(0.01\), and \(0.1\), while the  two rows
    show the \(Q=5~\mathrm{GeV}\) and \(Q=100~\mathrm{GeV}\) slices.}
    \label{fig:tmdpdf_kt_bands}
\end{figure}

The momentum-space TMD PDFs with uncertainty bands are shown in
fig.~\ref{fig:tmdpdf_kt_bands} for both the replica and Bayesian
extractions. In all cases, the extracted TMD PDFs are smooth and positive,
with a maximum at \(k_T=0\) and a monotonic falloff toward larger \(k_T\).
Comparing the \(Q=5~\mathrm{GeV}\) and \(Q=100~\mathrm{GeV}\) slices at fixed
\(x\), one observes the expected evolution-driven broadening: at the higher
scale, the distribution is suppressed in the low-\(k_T\) region and
relatively enhanced at larger \(k_T\), corresponding to a harder
transverse-momentum tail. This behavior is the momentum-space counterpart of
the stronger large-\(b\) suppression seen in the \(b\)-space distributions at
higher \(Q\). The \(x\)-dependence is also clearly visible. At \(Q=5~\mathrm{GeV}\), the
overall magnitude of the TMD PDFs increases with \(x\) over the plotted
range. By contrast, at \(Q=100~\mathrm{GeV}\) the \(x\)-dependence becomes
non-monotonic, with the magnitude first decreasing from $x=0.001$ to $x=0.01$, and then increasing as
\(x\) increases. The replica and Bayesian extractions preserve these
qualitative features and yield very similar central behavior. Their
differences are most noticeable in the low-\(k_T\) peak region, where the
Bayesian uncertainty bands are generally broader, while the two bands become
closer in the larger-\(k_T\) tail.

\subsection{Comparison to Data}

\begin{figure}[t]
    \centering
    \makebox[\textwidth][c]{\includegraphics[width=1.0\textwidth]{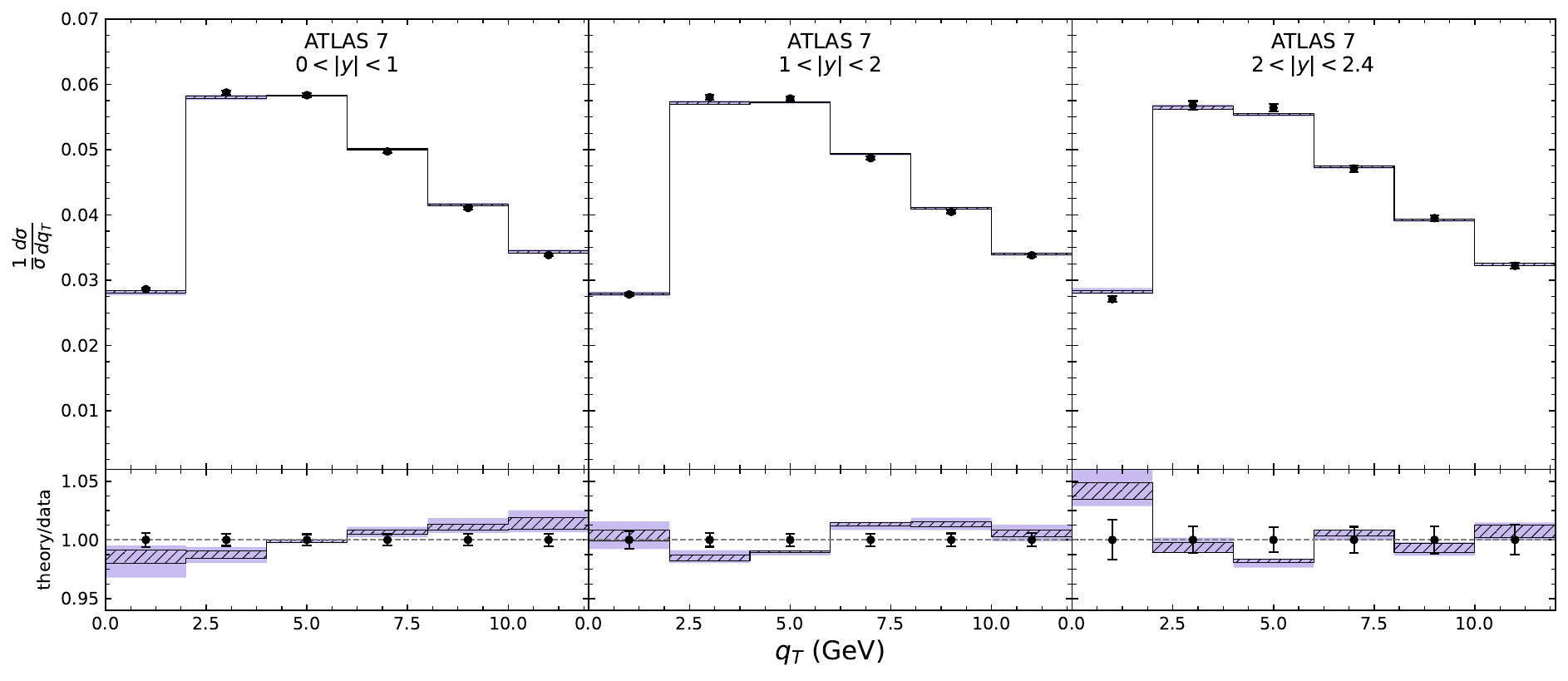}}
    \caption{Comparison of replica and Bayesian 68\% uncertainty bands
    for normalized ATLAS measurements at $\sqrt{s}=7~\mathrm{TeV}$.
    The purple bands show the Bayesian results, while the hatched bands show the replica results.}
    \label{fig:atlas7_replica_bayesian}
\end{figure}

\begin{figure}[t]
    \centering
    \makebox[\textwidth][c]{\includegraphics[width=1.0\textwidth]{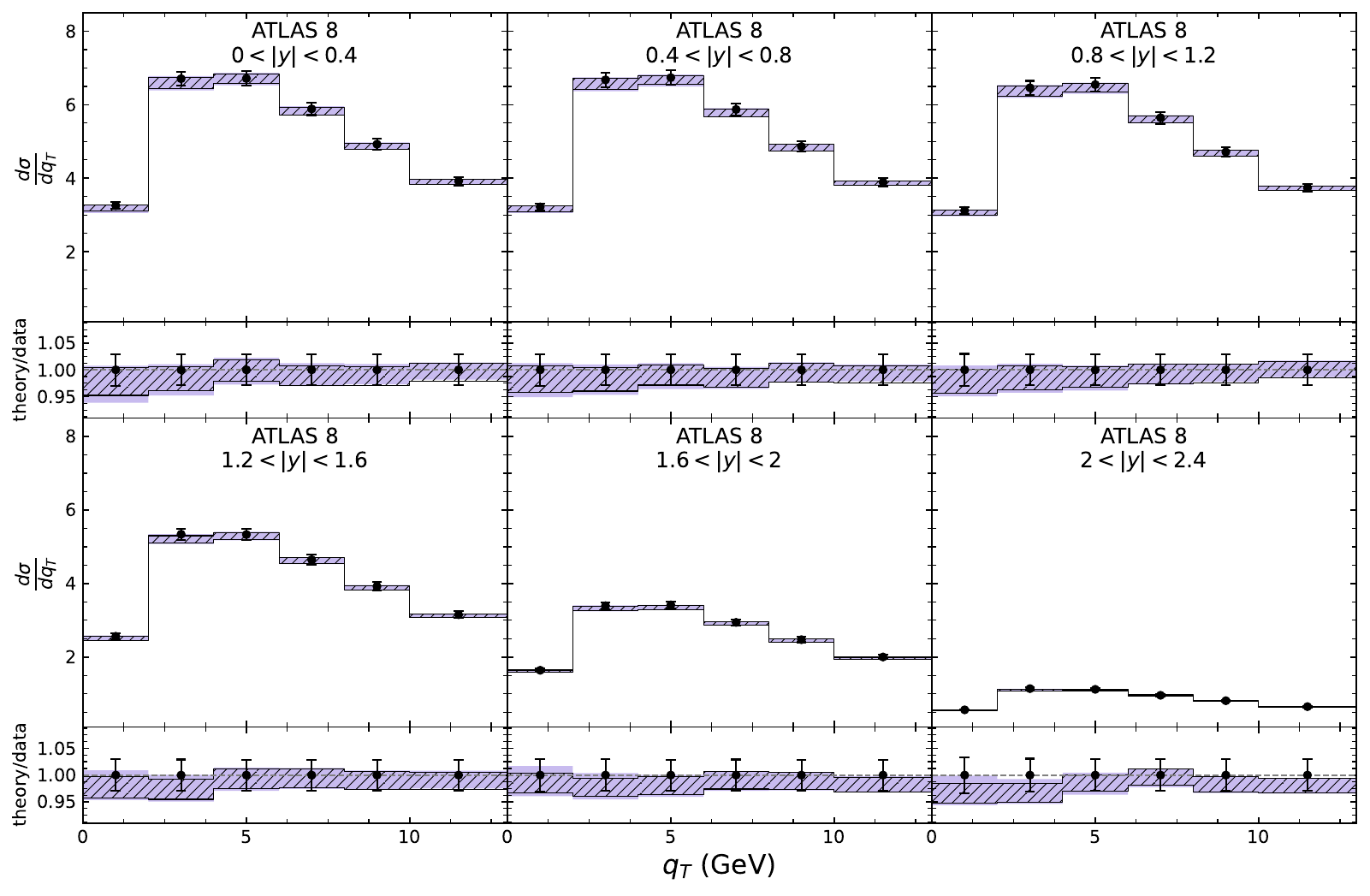}}
    \caption{Comparison of replica and Bayesian $68\%$ prediction intervals
    for normalized ATLAS measurements at $\sqrt{s}=8~\mathrm{TeV}$.}
    \label{fig:atlas8_replica_bayesian}
\end{figure}

\begin{figure}[t]
    \centering
    \includegraphics[width=0.8\textwidth]{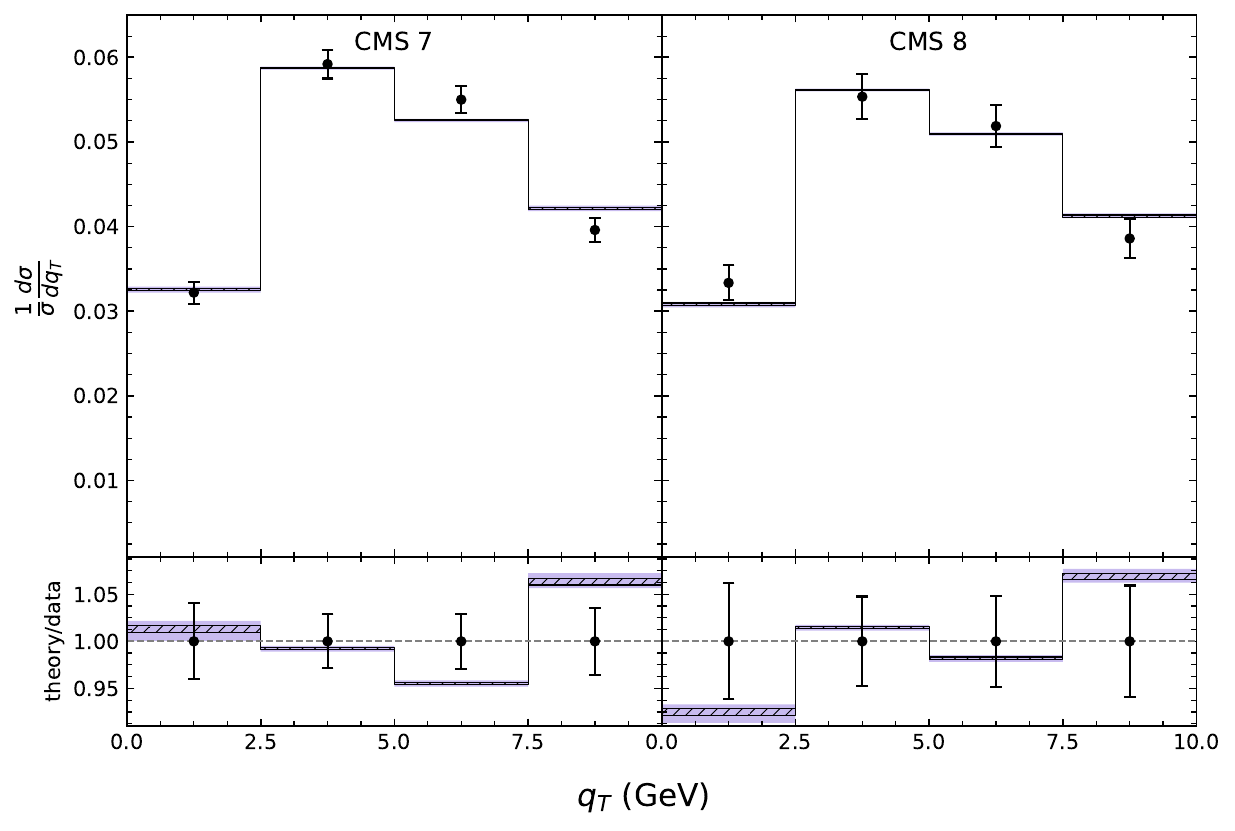}
    \caption{Replica and Bayesian uncertainty bands for CMS measurements.}
    \label{fig:cms_replica_bayesian}
\end{figure}

\begin{figure}[t]
    \centering
    \makebox[\textwidth][c]{\includegraphics[width=1.05\textwidth]{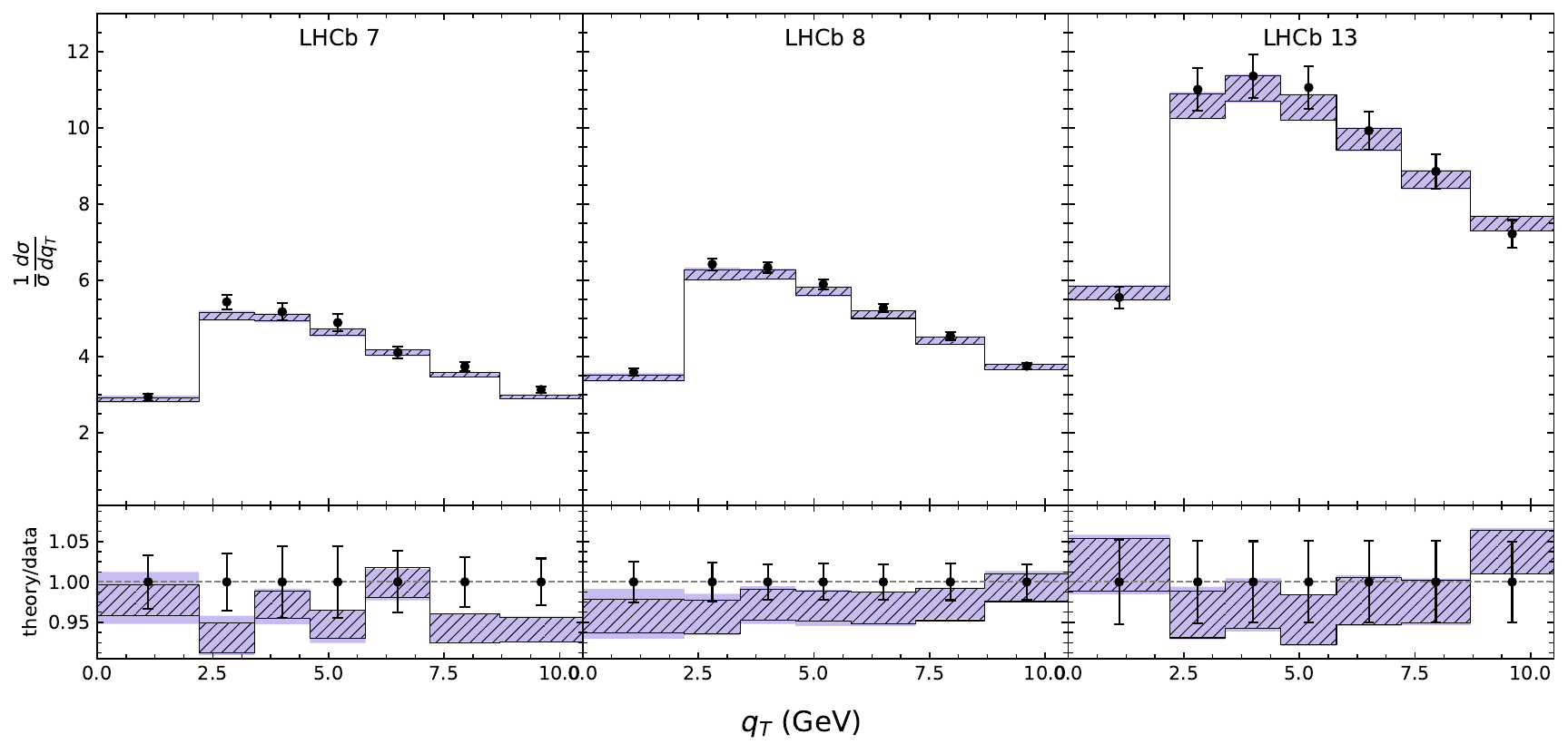}}
    \caption{Replica and Bayesian uncertainty bands for LHCb
    measurements.}
    \label{fig:lhcb_replica_bayesian}
\end{figure}

\begin{figure}[t]
    \centering
    \makebox[\textwidth][c]{\includegraphics[width=1.05\textwidth]{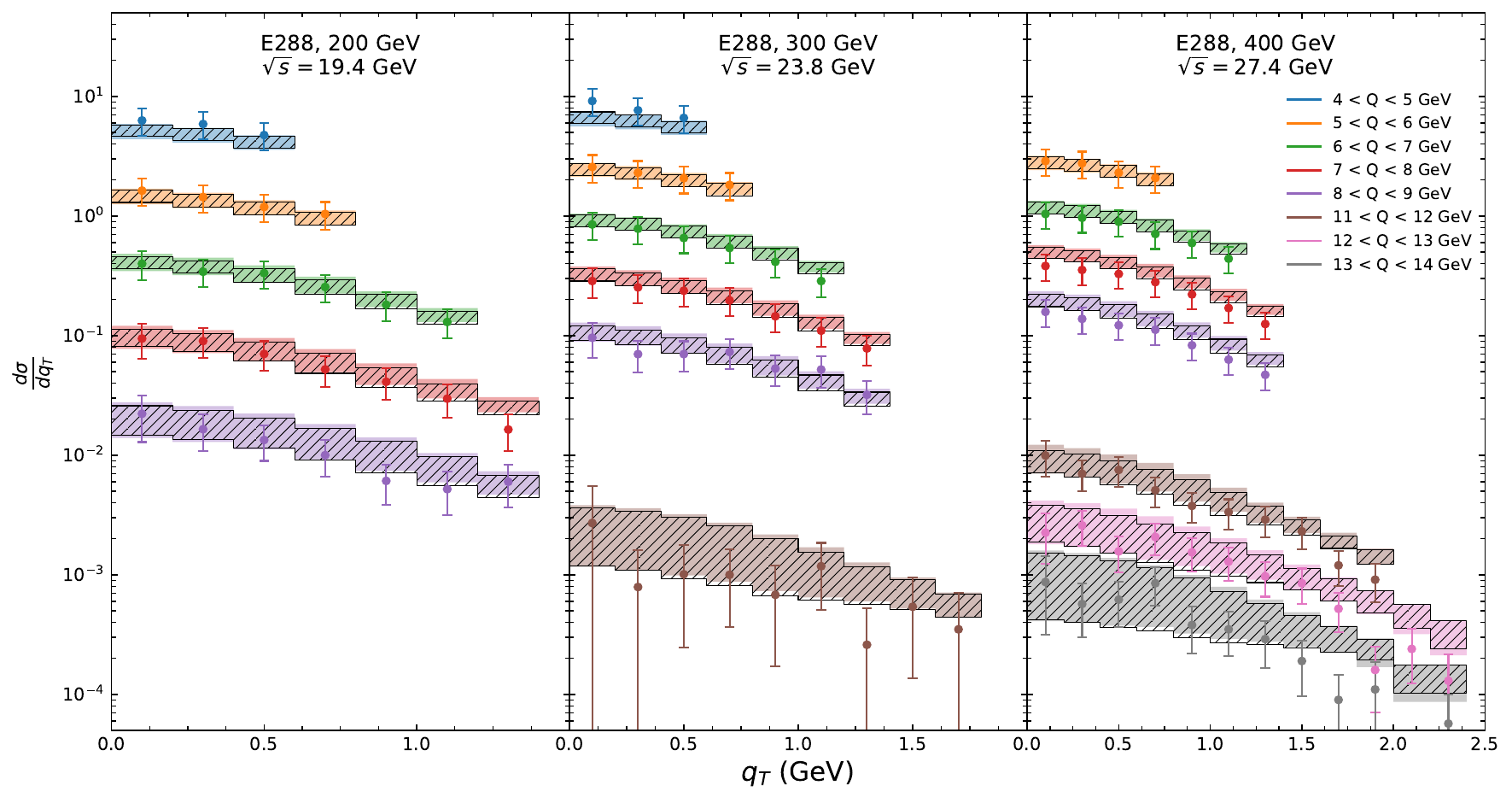}}
    \caption{Replica and Bayesian uncertainty comparison for the E288 fixed-target
    data set. The predictions are shown on a logarithmic scale.}
    \label{fig:e288_replica_bayesian}
\end{figure}

\begin{figure}[t]
    \centering
    \begin{minipage}[t]{0.49\textwidth}
        \centering
        \vspace{0pt}
        \includegraphics[width=\textwidth]{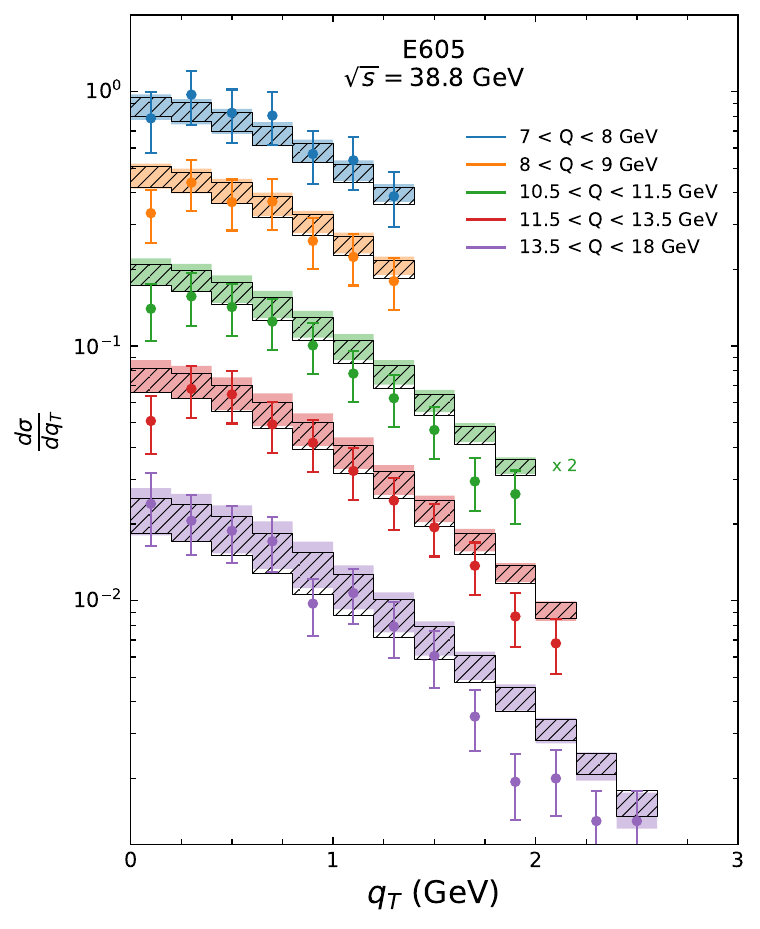}
    \end{minipage}\hfill
    \begin{minipage}[t]{0.49\textwidth}
        \centering
        \vspace{0pt}
        \includegraphics[width=\textwidth]{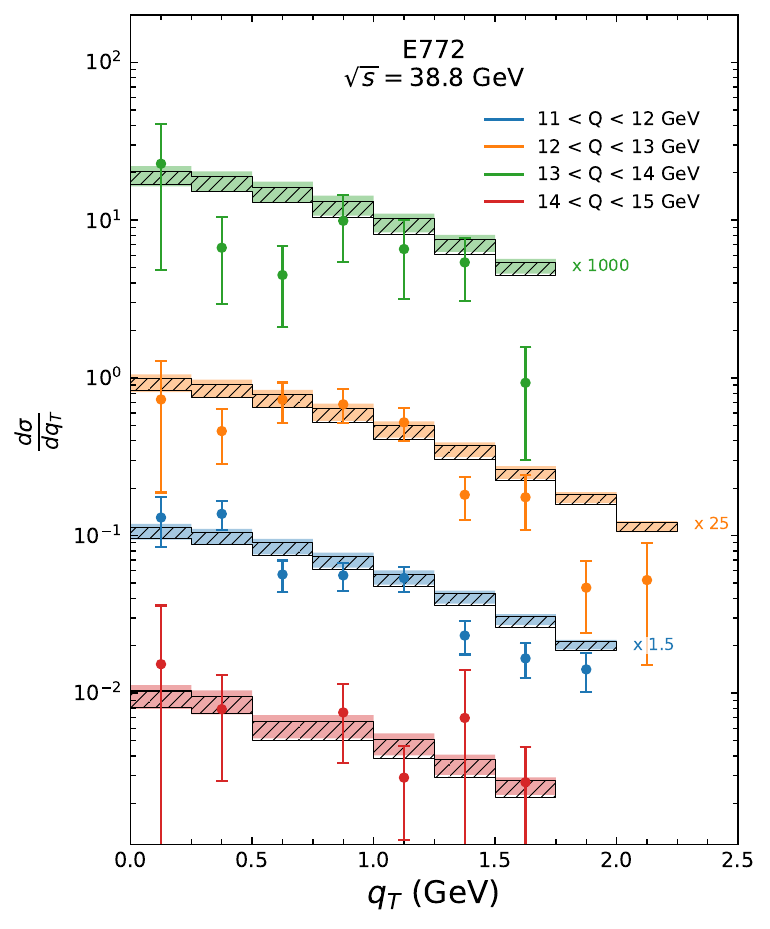}
    \end{minipage}
    \caption{Replica and Bayesian uncertainty comparison for the E605 (left) and
    E772 (right) fixed-target data sets. In both panels, some mass
    bins are displayed with the multiplicative factors indicated in the plot.}
    \label{fig:e605_e772_replica_bayesian}
\end{figure}

A representative comparison between theory predictions and data is shown in
figs.~\ref{fig:atlas7_replica_bayesian}--\ref{fig:e605_e772_replica_bayesian}.
Across both collider and fixed-target measurements, the central predictions from
the replica and Bayesian analyses are highly consistent, with the difference
between the two methods appearing primarily in the size of the uncertainty
bands rather than in the central curves themselves. For the ATLAS measurements
in figs.~\ref{fig:atlas7_replica_bayesian}--\ref{fig:atlas8_replica_bayesian},
the two central predictions are nearly indistinguishable over the full plotted
\(q_T\) range, as is most clearly seen in the lower theory/data panels, where
both remain close to unity in almost all bins. A similar pattern is observed in
the CMS and LHCb bins, while for the fixed-target observables in
figs.~\ref{fig:e288_replica_bayesian} and
\ref{fig:e605_e772_replica_bayesian} the two constructions remain comparably
close over most of the fitted range. The main visible effect of the Bayesian
analysis is therefore a modest broadening of the prediction bands.

This qualitative picture is supported by a direct comparison of the band
widths. Averaged over all 57 fitted datasets and 465 bins, the Bayesian
uncertainty bands are wider than the replica bands by a factor of 1.23. The
only datasets for which the mean Bayesian-to-replica band-width ratio exceeds 2
are ATLAS 7 TeV with \(|y|<1\), ATLAS 7 TeV with \(1<|y|<2\), CMS 7 TeV, and
CMS 8 TeV. These are all normalized collider measurements with very small
relative uncertainties, so even a sizable relative increase corresponds to a
small absolute change in the prediction band. Overall, the prediction plots show
that the two inference strategies yield highly consistent central cross
sections, with the Bayesian treatment providing a slightly more
conservative estimate of the uncertainties.

\subsection{Comparison of Bayesian and replica uncertainty estimates}
\label{subsec:replica-bayes-comparison}

At the parameter level, the differences between replica and Bayesian uncertainties can be clearly seen in 
the Gaussian limit. Such differences have been studied in the context of Collinear PDFs for linear prediction models \cite{DelDebbio:2022bayesClosure}, as well as more generally \cite{Watt:coll_pdf_hessian_2012, Costantini:2024replica}. Here we employ a very direct study of the loss curvature to frame our discussion of the differences between replica and Bayesian methods. For more detailed derivations and reference material, see Appendix \ref{sec:appA}.

For the replica method, the covariance is determined by the Fisher Information evaluated at the true parameter set $\theta_\ast$,
\begin{equation}
\mathrm{Cov}_{\rm rep}(\theta)
\approx
I(\theta_\ast)^{-1},
\label{eq:cov_rep_fisher}
\end{equation}
which is just the $\chi^{2}$ Hessian averaged over data and collinear PDF resamplings
\begin{equation}
I(\theta_{\ast})
=
\left\langle
H(\theta_\ast)
\right\rangle_{\mathrm{data},\,\mathrm{collinear}} \, .
\end{equation}

For the Bayesian method, the covariance is determined by the effective Hessian evaluated at the posterior maximizing $\theta_{\rm MAP}$,
\begin{equation}
\mathrm{Cov}_{\rm Bayes}(\theta)
\approx
H_{\rm eff}(\theta_{\rm MAP})^{-1},
\label{eq:cov_bayes_heff}
\end{equation}
where $H_{\rm eff}$ contasins both likelihood and prior contributions.
\begin{equation}
H_{\rm eff}(\theta)
=
H_{\rm like}(\theta)
+
H_{\rm prior}(\theta),
\qquad
\left\{
\begin{aligned}
H_{\rm like}(\theta)
&=
-\nabla_\theta \nabla_\theta \log \mathcal{L}(\theta),
\\[4pt]
H_{\rm prior}(\theta)
&=
-\nabla_\theta \nabla_\theta \log \pi(\theta).
\end{aligned}
\right.
\end{equation}

Because the fit geometries for bayesian and replica methodologies are both based on the same experimental data and choice of $\chi^{2}$, one might expect the two resulting uncertainties to be related. To make such a comparison, it is useful to organize both covariances
about the same central fit geometry. Let
\begin{equation}
L_{\rm cent}(\theta)\equiv -\log p(\mathcal D\mid \theta,\phi_\ast);
\qquad
\theta_{\rm cent}\equiv \arg\min_\theta L_{\rm cent}(\theta),
\end{equation}
and define the corresponding central Hessian
\begin{equation}
H_{\rm cent}
\equiv
\nabla_\theta^2 L_{\rm cent}(\theta)\big|_{\theta=\theta_{\rm cent}}.
\end{equation}
which is the object used to generate uncertainties using a standard Hessian Method \cite{Watt:coll_pdf_hessian_2012}. Then the replica and Bayesian inverse covariances may be written as corrections to this central fit geometry
\begin{equation}
\begin{aligned}
\mathrm{Cov}_{\rm rep}^{-1}(\theta)
\;\approx\;
I(\theta_\ast)
&\approx
H_{\rm cent}
+
\Delta H_{\rm rep},
\\[6pt]
\mathrm{Cov}_{\rm Bayes}^{-1}(\theta)
\;\approx\;
H_{\rm eff}(\theta_{\rm MAP})
&\approx
H_{\rm cent}
+
\Delta H_{\rm marg}
+
H_{\rm prior}
+
H_{\rm shift}.
\end{aligned}
\label{eq:rep_bayes_hcent_expand}
\end{equation}
where the correction terms can be found in Appendix \ref{sec:appA}. Each correction term corresponds to a particular aspect of the fitting methodology. The replica method has the simplest structure, and so receives one correction due to averaging result over replicas. The Bayesian structure is more rich--- we receive contributions due to the prior involvement $H_{\rm prior}$, our marginalization of the likelihood over collinear parameters $\Delta H_{\rm marg}$, and a shift contribution $H_{\rm shift}$ from the fact that $\theta_{\rm MAP}$ generally differs from $\theta_{\rm cent}$ (see Appendix \ref{sec:appA}).

Thus, the Gaussian limit eq.~\eqref{eq:rep_bayes_hcent_expand} indicates that we should expect different results from the replica vs Bayesian method. These differences and their agreement with the Hessian approximation can be seen in Table \ref{tab:stddev_bayes_replica_compare_physical_avg}. We find that the Hessian agreement is stronger for the Bayesian results than replica. Furthermore, the uncertainty differences can be described as dividing the fit parameters into two groups according to which
method yields the larger marginal uncertainty. The Bayesian-broader group is
\(\{\lambda_2,\lambda_3,\log x_0,\sigma_x\}\), while the replica-broader group
is \(\{\lambda_1,A,B_{\rm NP},c_0,c_1\}\). Among these, \(\sigma_x\) is the
most pronounced Bayesian-broader direction, whereas \(c_1\) and \(c_0\) are
the most pronounced replica-broader directions.

In addition to providing a useful basis for comparison, the Gaussian limits in Eqs.~\ref{eq:cov_rep_fisher} and \ref{eq:cov_bayes_heff} are also relevant for the use of fit results in subsequent fitting procedures and uncertainty propagation. In many applications, the choice of $\chi^{2}$, likelihood, or other cost function is itself based on a Gaussian approximation. For example, in this study we incorporate collinear PDF uncertainties through their contribution to the covariance matrix in eq.~\eqref{eq:collinear_pdf_covariance}, which is an appropriate description so long as those uncertainties are sufficiently close to the Gaussian limit. For our results, we find that the Gaussian/Hessian approximation for the marginal uncertainties is more accurate for the Bayesian fit than for the replica fit. This suggests that the Bayesian results may be better suited for use in fitting pipelines based on Gaussian-statistical loss functions. For the replica results, one may either account explicitly for non-Gaussian features of the fit distribution or, more simply, increase the number of replicas until the resulting distribution is adequately described by the Hessian approximation. In this sense, comparison to the Hessian limit may also serve as a practical diagnostic for determining how many replicas are needed in a replica-based fit. 

\begin{table}[t]
    \centering
    \begin{tabular}{lccccc}
        \toprule
        & \multicolumn{2}{c}{Bayesian} & \multicolumn{2}{c}{Replica} & \\
        \cmidrule(lr){2-3} \cmidrule(lr){4-5}
        Parameter & Posterior & $H_{\rm eff}(\theta_{\rm MAP})$ Approx. & Posterior & $I(\theta_{\rm cent})$ Approx. & Bayes / Replica \\
        \midrule
        $\lambda_1$       & 0.03400 & 0.03620 & 0.03750 & 0.01692 & 0.90667 \\
        $\lambda_2$       & 0.18500 & 0.19553 & 0.16000 & 0.10216 & 1.15625 \\
        $\lambda_3$       & 0.48000 & 0.56223 & 0.43500 & 0.33596 & 1.10345 \\
        $x_0$             & 0.00175 & 0.00170 & 0.00175 & 0.00290 & 1.00000 \\
        $\sigma_x$        & 0.34000 & 0.35951 & 0.12000 & 0.06480 & 2.83333 \\
        $A_{\mathrm{NP}}$ & 0.14000 & 0.13344 & 0.13000 & 0.35889 & 1.07692 \\
        $B_{\mathrm{NP}}$ & 0.14000 & 0.18445 & 0.13500 & 0.07876 & 1.03704 \\
        $c_0$             & 0.00335 & 0.00403 & 0.00585 & 0.00997 & 0.57265 \\
        $c_1$             & 0.00245 & 0.00315 & 0.00755 & 0.00401 & 0.32450 \\
        \bottomrule
    \end{tabular}
    \caption{Posterior marginal parameter uncertainties in physical parameter space. For the posterior columns, asymmetric uncertainties from Table~\ref{tab:fit_parameters} have been symmetrized by averaging the upper and lower errors. The Hessian-based columns show the corresponding Gaussian approximations.}
    \label{tab:stddev_bayes_replica_compare_physical_avg}
\end{table}

\section{Conclusions}
\label{sec:conclusions}

In this work, we have presented a global extraction of unpolarized quark TMD PDFs from Drell--Yan data within a Bayesian inference framework at N$^3$LO+N$^4$LL accuracy, and compared the results directly to a replica-based analysis in the same theoretical setup. We find that the two methods lead to consistent central fits, with overall fit quality at the level of $\chi^2/N \simeq 1$. The Bayesian analysis yields broader prediction bands for data points, while preserving the same qualitative behavior for the extracted nonperturbative parameters, Collins--Soper kernel, TMD PDFs, and cross sections. The extracted Collins--Soper kernel from this work is closest to the EEC extraction and are consistent with lattice determinations of ASWZ24 and LPC23, and lies between the ART23 and ART25 TMD extractions.

A central component of this work is the integration of AI-assisted tools into the TMD analysis workflow. We employed an AI-driven procedure to explore and rank candidate functional forms for the nonperturbative sector, enabling a broader and more systematic search of the ansatz space than is typically feasible with manual approaches. In addition, we constructed a machine-learning emulator for the TMD cross section, allowing for efficient likelihood evaluation and scalable Bayesian inference in a high-dimensional parameter space.

Within this framework, we performed a full posterior analysis of the nonperturbative parameters governing TMD PDFs and the Collins--Soper kernel. The resulting parameter distributions provide a direct probabilistic characterization of uncertainties and correlations. In parallel, we carried out a replica-based analysis within the same theoretical setup, enabling a controlled comparison between the two approaches. We find that the two methods yield broadly consistent central values, while differing in the structure of the inferred uncertainties, in particular in the resulting parameter correlations and the shapes of the parameter distributions. Our results demonstrate that Bayesian inference provides a complementary perspective on uncertainty quantification in TMD extractions, offering a transparent framework for exploring the structure of the parameter space and systematically incorporating additional sources of information.

Looking ahead, the framework developed in this work can be extended in several important directions. The Bayesian approach naturally accommodates the inclusion of heterogeneous inputs, such as lattice QCD constraints and future high-precision measurements, including those from the Electron--Ion Collider. In addition, the combination of AI-assisted modeling with probabilistic inference opens new possibilities for more flexible and less biased parametrizations of TMDs. These developments will play a key role in advancing precision studies of nucleon structure and in fully exploiting the physics potential of upcoming experiments.

\appendix

\section{Gaussian approximation and expansion about the central geometry}
\label{sec:appA}

Here we derive the Gaussian approximations underlying the replica
and Bayesian parameter covariances, and then reorganize both results about a
common central fit geometry. This provides the basis for the comparison used in
Sec.~\ref{subsec:replica-bayes-comparison}. Many of the items described here are familiar to the statistics community in the context of Gaussian/Laplace approximations and Fisher Information, and can be explored further here \cite{Efron:1994bootstrap, Stuart_2010, Kass1990TheVO}, for example. We employ a detailed derivation to make the contributions to replica and Bayesian covariances as clear as possible.  

\subsection{Replica covariance and the Fisher information}

For a given replica realization \(r\), let the fit objective be
\begin{equation}
\mathcal L_r(\theta) \equiv \frac12 \chi_r^2(\theta),
\end{equation}
and let \(\theta_r\) denote the corresponding best-fit point,
\begin{equation}
\nabla_\theta \mathcal L_r(\theta_r)=0.
\end{equation}
Expanding about the true parameter \(\theta_\ast\) gives
\begin{equation}
0
\approx
\nabla_\theta \mathcal L_r(\theta_\ast)
+
H_r(\theta_\ast)(\theta_r-\theta_\ast),
\end{equation}
so that
\begin{equation}
\theta_r-\theta_\ast
\approx
-\,H_r(\theta_\ast)^{-1}\nabla_\theta \mathcal L_r(\theta_\ast).
\label{eq:theta_r_score_expand_app}
\end{equation}
The replica covariance may therefore be written as
\begin{equation}
\mathrm{Cov}_{\rm rep}(\theta)
=
\Big\langle
(\theta_r-\theta_\ast)(\theta_r-\theta_\ast)^{\mathsf T}
\Big\rangle_{\rm data,\;collinear}
\approx
\Big\langle
H_r^{-1}
\bigl(\nabla_\theta \mathcal L_r\bigr)
\bigl(\nabla_\theta \mathcal L_r\bigr)^{\mathsf T}
H_r^{-1}
\Big\rangle_{\theta=\theta_\ast}.
\label{eq:cov_rep_score_form_app}
\end{equation}
In the large-sample Gaussian regime, the replica Hessian fluctuates only weakly
across the ensemble, so to leading order one may replace it by its average,
\begin{equation}
I(\theta_\ast)
\equiv
\Big\langle H_r(\theta_\ast)\Big\rangle_{\rm data,\;collinear}.
\label{eq:fisher_avg_def_app_reorg}
\end{equation}
Using the standard regularity relation between expected Hessian and score
covariance,
\begin{equation}
\Big\langle
\nabla_\theta \mathcal L_r(\theta_\ast)
\nabla_\theta \mathcal L_r(\theta_\ast)^{\mathsf T}
\Big\rangle_{\rm data,\;collinear}
\approx
I(\theta_\ast),
\label{eq:score_fisher_identity_app}
\end{equation}
one then finds
\begin{equation}
\mathrm{Cov}_{\rm rep}(\theta)
\approx
I(\theta_\ast)^{-1}
\,I(\theta_\ast)\,
I(\theta_\ast)^{-1}
=
I(\theta_\ast)^{-1}.
\label{eq:cov_rep_fisher_app_reorg}
\end{equation}
Thus, in the Gaussian approximation, the replica covariance is governed by the
inverse Fisher information.
Here \(\theta_\ast\) denotes the true parameter, or equivalently the point about
which the replica ensemble is centered in the large-sample Gaussian limit.

\subsection{Bayesian covariance and the effective Hessian}

For the Bayesian treatment, the posterior is
\begin{equation}
p(\theta\mid\mathcal D)
\propto
p(\mathcal D\mid\theta)\,\pi(\theta),
\label{eq:posterior_def_app_reorg}
\end{equation}
where \(\pi(\theta)\) is the prior. It is convenient to define the action
\begin{equation}
S(\theta)
\equiv
-\log p(\mathcal D\mid\theta)-\log \pi(\theta).
\label{eq:bayes_action_app_reorg}
\end{equation}
The maximum a posteriori point is then
\begin{equation}
\theta_{\rm MAP}
\equiv
\arg\min_\theta S(\theta)
=
\arg\max_\theta p(\theta\mid\mathcal D).
\label{eq:theta_map_def_app_reorg}
\end{equation}

Expanding the action about \(\theta_{\rm MAP}\) gives
\begin{equation}
S(\theta)
\approx
S(\theta_{\rm MAP})
+
\frac12(\theta-\theta_{\rm MAP})^{\mathsf T}
H_{\rm eff}(\theta_{\rm MAP})
(\theta-\theta_{\rm MAP}),
\label{eq:action_quad_expand_app_reorg}
\end{equation}
with effective Hessian
\begin{equation}
H_{\rm eff}(\theta)
=
H_{\rm like}(\theta)+H_{\rm prior}(\theta),
\label{eq:heff_def_app_reorg}
\end{equation}
where
\begin{equation}
H_{\rm like}(\theta)
=
-\nabla_\theta\nabla_\theta \log p(\mathcal D\mid\theta),
\qquad
H_{\rm prior}(\theta)
=
-\nabla_\theta\nabla_\theta \log \pi(\theta).
\label{eq:h_like_h_prior_app_reorg}
\end{equation}
In the Gaussian regime, the Bayesian covariance is therefore
\begin{equation}
\mathrm{Cov}_{\rm Bayes}(\theta)
\approx
H_{\rm eff}(\theta_{\rm MAP})^{-1}.
\label{eq:cov_bayes_heff_app_reorg}
\end{equation}

The likelihood appearing here is itself marginalized over collinear-PDF
parameters,
\begin{equation}
p(\mathcal D\mid\theta)
=
\int d\phi\;
p(\mathcal D\mid\theta,\phi)\,p(\phi),
\label{eq:pdf_marg_like_formal_app_reorg}
\end{equation}
which in practice is approximated by an average over collinear-PDF replicas,
\begin{equation}
p(\mathcal D\mid\theta)
\approx
\frac{1}{N_\phi}\sum_{r=1}^{N_\phi}
p(\mathcal D\mid\theta,\phi_r).
\label{eq:pdf_marg_like_rep_app_reorg}
\end{equation}
Unlike the replica Fisher information, this average is taken at the likelihood
level rather than at the Hessian level. Because the logarithm is taken only
after the replica average, the resulting likelihood curvature is not simply the
average of fixed-replica Hessians. This is the basic reason the Bayesian and
replica curvature structures need not coincide, even before the prior is added.

\subsection{Expansion about the central fit geometry}

We now reorganize both inverse covariances about a common central geometry. Let the central negative log-likelihood be defined using the experimental
central values and the central collinear PDF,
\begin{equation}
L_{\rm cent}(\theta)
\equiv
-\log p(\mathcal D\mid\theta,\phi_\ast),
\label{eq:Lcent_def_app_reorg}
\end{equation}
and let
\begin{equation}
\theta_{\rm cent}
\equiv
\arg\min_\theta L_{\rm cent}(\theta)
\label{eq:thetacent_def_app_reorg}
\end{equation}
denote the corresponding central best-fit point. The associated central Hessian
is
\begin{equation}
H_{\rm cent}
\equiv
\nabla_\theta^2 L_{\rm cent}(\theta)\big|_{\theta=\theta_{\rm cent}}.
\label{eq:Hcent_def_app_reorg}
\end{equation}
By construction,
\begin{equation}
\nabla_\theta L_{\rm cent}(\theta_{\rm cent})=0.
\label{eq:gradLcent_zero_app_reorg}
\end{equation}
For each replica \(r\), write the replica objective as
\begin{equation}
\mathcal L_r(\theta)
=
L_{\rm cent}(\theta)+\delta L_r(\theta),
\label{eq:Lr_split_app_reorg}
\end{equation}
where \(\delta L_r\) encodes the perturbation due to data and collinear-PDF
resampling. The replica best-fit point satisfies
\begin{equation}
\nabla_\theta \mathcal L_r(\theta_r)=0.
\end{equation}
Expanding around \(\theta_{\rm cent}\) yields
\begin{equation}
0
\approx
\nabla_\theta \delta L_r(\theta_{\rm cent})
+
H_{\rm cent}(\theta_r-\theta_{\rm cent}),
\end{equation}
so that
\begin{equation}
\theta_r-\theta_{\rm cent}
\approx
-\,H_{\rm cent}^{-1}\nabla_\theta \delta L_r(\theta_{\rm cent}).
\label{eq:theta_r_shift_app_reorg}
\end{equation}
If the resampling is unbiased, then
\begin{equation}
\Big\langle \nabla_\theta \delta L_r(\theta_{\rm cent}) \Big\rangle_{\rm data,\;collinear}=0,
\end{equation}
and therefore
\begin{equation}
\langle \theta_r \rangle \approx \theta_{\rm cent}.
\label{eq:theta_rep_mean_app_reorg}
\end{equation}

Next, evaluate the replica Hessian average at the central point:
\begin{equation}
I(\theta_{\rm cent})
=
\Big\langle H_r(\theta_{\rm cent})\Big\rangle_{\rm data,\;collinear}.
\end{equation}
This may be written as
\begin{equation}
I(\theta_{\rm cent})
=
H_{\rm cent}
+
\Delta H_{\rm rep},
\label{eq:fisher_hcent_expand_app_reorg}
\end{equation}
with
\begin{equation}
\Delta H_{\rm rep}
\equiv
\Big\langle H_r(\theta_{\rm cent})\Big\rangle_{\rm data,\;collinear}
-
H_{\rm cent}.
\label{eq:DeltaH_rep_def_app_reorg}
\end{equation}
Accordingly,
\begin{equation}
\mathrm{Cov}_{\rm rep}(\theta)
\approx
\left(H_{\rm cent}+\Delta H_{\rm rep}\right)^{-1}.
\label{eq:cov_rep_hcent_expand_app_reorg}
\end{equation}

To compare the Bayesian fit point to the central one, write the marginalized
Bayesian likelihood as a correction to the central likelihood. Define
\begin{equation}
p(\mathcal D\mid\theta,\phi_r)
=
p(\mathcal D\mid\theta,\phi_\ast)+\delta p_r(\theta),
\end{equation}
so that
\begin{equation}
p(\mathcal D\mid\theta)
\approx
p(\mathcal D\mid\theta,\phi_\ast)
+
\frac1{N_\phi}\sum_{r=1}^{N_\phi}\delta p_r(\theta).
\end{equation}
Then
\begin{equation}
-\log p(\mathcal D\mid\theta)
=
L_{\rm cent}(\theta)
-
\log\!\left[
1+
\frac1{N_\phi}\sum_{r=1}^{N_\phi}
\frac{\delta p_r(\theta)}{p(\mathcal D\mid\theta,\phi_\ast)}
\right].
\end{equation}
For small collinear-PDF-induced deviations, this becomes
\begin{equation}
-\log p(\mathcal D\mid\theta)
\approx
L_{\rm cent}(\theta)
+
\Delta L_{\rm marg}(\theta),
\label{eq:DeltaLmarg_app_reorg}
\end{equation}
with
\begin{equation}
\Delta L_{\rm marg}(\theta)
\equiv
-
\frac1{N_\phi}\sum_{r=1}^{N_\phi}
\frac{\delta p_r(\theta)}{p(\mathcal D\mid\theta,\phi_\ast)}.
\label{eq:DeltaLmarg_def_app_reorg}
\end{equation}
The posterior action may therefore be written as
\begin{equation}
S(\theta)
=
L_{\rm cent}(\theta)
+
\Delta L_{\rm marg}(\theta)
-
\log \pi(\theta).
\label{eq:S_split_central_app_reorg}
\end{equation}
Using the MAP condition \(\nabla_\theta S(\theta_{\rm MAP})=0\) and expanding
around \(\theta_{\rm cent}\), one finds
\begin{equation}
\theta_{\rm MAP}-\theta_{\rm cent}
\approx
-\,H_{\rm cent}^{-1}
\left[
\nabla_\theta \Delta L_{\rm marg}(\theta_{\rm cent})
-
\nabla_\theta \log \pi(\theta_{\rm cent})
\right].
\label{eq:thetaMAP_shift_full_app_reorg}
\end{equation}

We now expand the Bayesian effective Hessian about \(\theta_{\rm cent}\). Since
the marginalized likelihood differs from the central likelihood by
\(\Delta L_{\rm marg}\), its Hessian at the central point is
\begin{equation}
H_{\rm like}(\theta_{\rm cent})
=
H_{\rm cent}
+
\Delta H_{\rm marg},
\label{eq:Hlike_central_expand_app_reorg}
\end{equation}
where
\begin{equation}
\Delta H_{\rm marg}
\equiv
\nabla_\theta^2 \Delta L_{\rm marg}(\theta)\big|_{\theta=\theta_{\rm cent}}.
\label{eq:DeltaHmarg_def_app_reorg}
\end{equation}
Thus \(\Delta H_{\rm marg}\) is the leading curvature correction generated by
performing the collinear-PDF average at the likelihood level rather than
working directly with the fixed-\(\phi_\ast\) central likelihood.

Next expand the effective Hessian from \(\theta_{\rm cent}\) to
\(\theta_{\rm MAP}\):
\begin{equation}
H_{\rm eff}(\theta_{\rm MAP})
\approx
H_{\rm eff}(\theta_{\rm cent})
+
\sum_c (\theta_{\rm MAP}-\theta_{\rm cent})_c\,
\partial_c H_{\rm eff}(\theta)\big|_{\theta=\theta_{\rm cent}}.
\label{eq:Heff_shift_expand_app_reorg}
\end{equation}
Using
\begin{equation}
H_{\rm eff}(\theta_{\rm cent})
=
H_{\rm cent}
+
\Delta H_{\rm marg}
+
H_{\rm prior},
\qquad
H_{\rm prior}
\equiv
-\nabla_\theta\nabla_\theta \log\pi(\theta)\big|_{\theta=\theta_{\rm cent}},
\end{equation}
we obtain
\begin{equation}
H_{\rm eff}(\theta_{\rm MAP})
\approx
H_{\rm cent}
+
\Delta H_{\rm marg}
+
H_{\rm prior}
+
H_{\rm shift},
\label{eq:heff_hcent_expand_app_reorg}
\end{equation}
where
\begin{equation}
H_{\rm shift}
\equiv
\sum_c (\theta_{\rm MAP}-\theta_{\rm cent})_c\,
\partial_c H_{\rm eff}(\theta)\big|_{\theta=\theta_{\rm cent}}
\end{equation}
is the leading correction induced by evaluating the effective Hessian at
\(\theta_{\rm MAP}\) rather than at \(\theta_{\rm cent}\). Thus the Bayesian covariance becomes
\begin{equation}
\mathrm{Cov}_{\rm Bayes}(\theta)
\approx
\left(
H_{\rm cent}
+
\Delta H_{\rm marg}
+
H_{\rm prior}
+
H_{\rm shift}
\right)^{-1}.
\label{eq:cov_bayes_hcent_expand_app_reorg}
\end{equation}
We therefore arrive at the common expansion structure used in
Sec.~\ref{subsec:replica-bayes-comparison}:
\begin{equation}
\mathrm{Cov}_{\rm rep}^{-1}(\theta)
\approx
I(\theta_\ast)
\approx
H_{\rm cent}
+
\Delta H_{\rm rep},
\label{eq:rep_final_expand_app_reorg}
\end{equation}
and
\begin{equation}
\mathrm{Cov}_{\rm Bayes}^{-1}(\theta)
\approx
H_{\rm eff}(\theta_{\rm MAP})
\approx
H_{\rm cent}
+
\Delta H_{\rm marg}
+
H_{\rm prior}
+
H_{\rm shift}.
\label{eq:bayes_final_expand_app_reorg}
\end{equation}

The meaning of the various terms is then transparent:
\(H_{\rm cent}\) is the common central fit geometry,
\(\Delta H_{\rm rep}\) is the correction induced by replica averaging over
perturbed geometries,
\(\Delta H_{\rm marg}\) is the extra curvature generated by marginalizing over
collinear-PDF replicas,
\(H_{\rm prior}\) is the direct curvature from the prior,
and \(H_{\rm shift}\) is the leading correction associated with the displacement
of \(\theta_{\rm MAP}\) away from \(\theta_{\rm cent}\).

\section{Posterior Distributions}

\begin{figure}[t]
    \centering
    \makebox[\textwidth][c]{\includegraphics[width=1.0\textwidth]{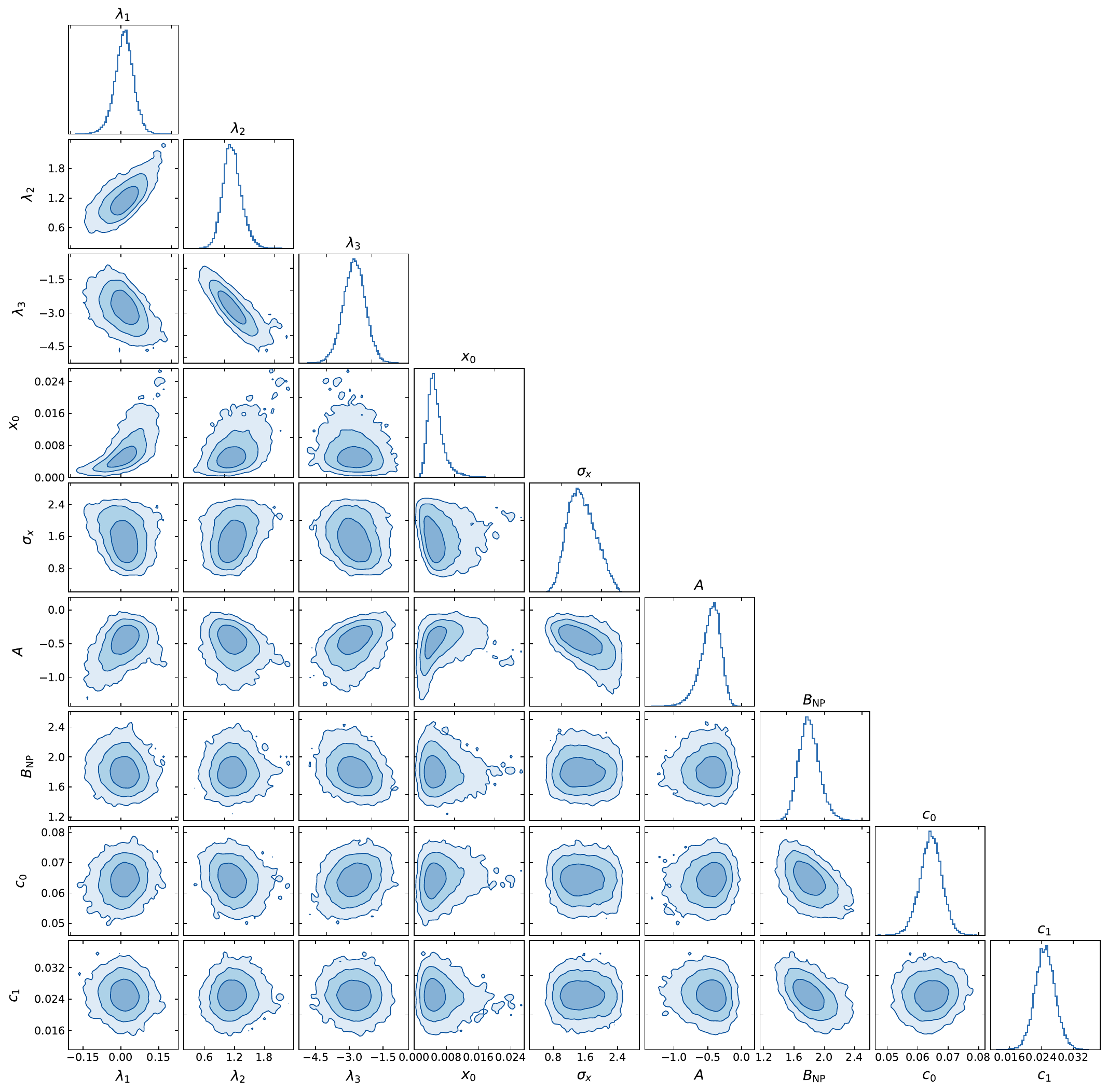}}
    \caption{Pair plot for the Bayesian posterior distribution of the 9 nonperturbative fit parameters. The diagonal panels show the one-dimensional marginalized posterior distributions, while the off-diagonal panels display the corresponding two-dimensional marginalized distributions.}
    \label{fig:cornerplot}
\end{figure}

In Fig.~\ref{fig:cornerplot}, we show the pair plot of the Bayesian posterior distribution for the nonperturbative parameters. The diagonal panels display the one-dimensional marginalized posterior distributions, while the off-diagonal panels show the corresponding two-dimensional projections. The distributions are generally well localized, and several parameter pairs exhibit visible correlations, consistent with the correlation matrix presented in the main text. This representation provides a detailed visualization of the posterior structure in parameter space.

\section*{Acknowledgments}
This work is supported by the National Science Foundation under grant No.~PHY-2515057. Curtis Zhou is also supported by the REU program at the UCLA Department of Physics and Astronomy. We thank Matteo Cerutti for helpful correspondence clarifying the treatment and propagation of collinear PDF uncertainties.

\bibliographystyle{JHEP-2modlong.bst}
\bibliography{references}

\end{document}